%% file: IEEE-conference-template-062824.tex
\documentclass[conference]{IEEEtran}
\IEEEoverridecommandlockouts
\usepackage{cite}
\usepackage{amsmath,amssymb,amsfonts}
\usepackage{algorithmic}
\usepackage{graphicx}
\usepackage{textcomp}
\usepackage{xcolor}
\usepackage{booktabs}
\usepackage{tabularx}
\usepackage{makecell}
\usepackage{array}
\usepackage{enumitem}
\usepackage{subcaption}
\usepackage{xcolor}
\usepackage[normalem]{ulem} % use normalem to keep \emph as italic

\newif\ifshowoldtext
\showoldtextfalse % hide removed text

\newcommand{\oldtext}[1]{%
  \ifshowoldtext
    \textcolor{red}{\sout{#1}}%
  \fi
}

\newcommand{\newtext}[1]{\textcolor{black}{#1}}

\usepackage{listings}
\definecolor{codegreen}{rgb}{0,0.6,0}
\definecolor{codegray}{rgb}{0.5,0.5,0.5}
\definecolor{codepurple}{rgb}{0.58,0,0.82}
\definecolor{backcolour}{rgb}{0.95,0.95,0.92}

\lstdefinestyle{mystyle}{
    backgroundcolor=\color{backcolour},   
    commentstyle=\color{codegreen},
    keywordstyle=\color{magenta},
    numberstyle=\tiny\color{codegray},
    stringstyle=\color{codepurple},
    basicstyle=\ttfamily\footnotesize,
    breakatwhitespace=false,         
    breaklines=true,                 
    captionpos=b,                    
    keepspaces=true,                 
    numbers=left,                    
    numbersep=5pt,                  
    showspaces=false,                
    showstringspaces=false,
    showtabs=false,                  
    tabsize=2
}

\usepackage{hyperref}
\newcolumntype{Y}{>{\raggedright\arraybackslash}X}

\def\BibTeX{{\rm B\kern-.05em{\sc i\kern-.025em b}\kern-.08em
    T\kern-.1667em\lower.7ex\hbox{E}\kern-.125emX}}
\begin{document}

\title{Language-Guided Multimodal Texture Authoring via Generative Models\\
% {\footnotesize \textsuperscript{*}Note: Sub-titles are not captured for https://ieeexplore.ieee.org  and
\thanks{This research
was supported by the National Science Foundation under
Grant No. 2504241.}
}

% \author{\IEEEauthorblockN{Anonymous Author}
% \and
% \IEEEauthorblockN{Anonymous Author}
% \and
% \IEEEauthorblockN{Anonymous Author}
% \and
% \IEEEauthorblockN{Anonymous Author}
% \and
% \IEEEauthorblockN{Anonymous Author}

% }

\makeatletter
\newcommand{\linebreakand}{%
  \end{@IEEEauthorhalign}
  \hfill\mbox{}\par
  \mbox{}\hfill\begin{@IEEEauthorhalign}
}
\makeatother

\author{\IEEEauthorblockN{Wanli Qian}
\IEEEauthorblockA{\textit{Department of Computer Science} \\
\textit{University of Southern California}\\
Los Angeles, USA \\
wanliqia@usc.edu}
\and
\IEEEauthorblockN{Aiden Chang}
\IEEEauthorblockA{\textit{Department of Computer Science} \\
\textit{University of Southern California}\\
Los Angeles, USA  \\
aidencha@usc.edu}
\and
\IEEEauthorblockN{Shihan Lu}
\IEEEauthorblockA{\textit{Center for Robotics and Biosystems} \\
\textit{Northwestern University}\\
Evanston, USA \\
shihanlu@northwestern.edu}
% \and
\linebreakand
\IEEEauthorblockN{Michael Gu}
\IEEEauthorblockA{\textit{Department of Computer Science} \\
\textit{University of Southern California}\\
Los Angeles, USA  \\
mxgu@usc.edu}
\and
\IEEEauthorblockN{Heather Culbertson}
\IEEEauthorblockA{\textit{Department of Computer Science} \\
\textit{University of Southern California}\\
Los Angeles, USA  \\
hculbert@usc.edu}

}

\maketitle
 
\begin{abstract}
Authoring realistic haptic textures typically requires low-level parameter tuning and repeated trial-and-error, limiting speed, transparency, and creative reach. We present a language-driven authoring system that turns natural-language prompts into multimodal textures: two coordinated haptic channels—sliding vibrations via force/speed-conditioned autoregressive (AR) models and tapping transients—and a text-prompted visual preview from a diffusion model. A shared, language-aligned latent links modalities so a single prompt yields semantically consistent haptic and visual signals; designers can write goals (e.g., ``gritty but cushioned surface,'' ``smooth and hard metal surface'') and immediately see and feel the result through a 3D haptic device. To verify that the learned latent encodes perceptually meaningful structure, we conduct an anchor-referenced, attribute-wise evaluation for roughness, slipperiness, and hardness. Participant ratings are projected to the interpretable line between two real-material references, revealing consistent trends—asperity effects in roughness, compliance in hardness, and surface-film influence in slipperiness. A human-subject study further indicates coherent cross-modal experience and low effort for prompt-based iteration. The results show that language can serve as a practical control modality for texture authoring: prompts reliably steer material semantics across haptic and visual channels, enabling a prompt-first, designer-oriented workflow that replaces manual parameter tuning with interpretable, text-guided refinement.

\end{abstract}


\input{sections/1-Introduction}
\input{sections/2-related}

\input{sections/3-method}

\input{sections/4-result}

\input{sections/5-usrEval}
\input{sections/6-Discussion}

\input{sections/7-conclusion}
\bibliographystyle{IEEEtran}
\newpage
\bibliography{ref}

\end{document}

%% file: sections/1-Introduction.tex
\section{Introduction}

% \textcolor{red}{SL: I feel that we need to bring up the concept of ``Texture Authoring'' earlier and more clearly. It may be more natural to center the introduction section around the texture authoring, instead of the methodology of texture modeling and rendering. Texture authoring can be a concept raised from the ``dataset-bounded'' limitations in data-driven methods, or just due to the needs of free design for virtual textures, so researchers explored different approaches, such as preference-driven, etc.
% }

% \textcolor{red}{SL: Additionally, ``Texture Authoring'' is not a brand new concept~\cite{8710002}. Researchers brought it up before, such as based on affective space, but mainly on single-modality or using linear interpolation methods. As the title said, I guess another major motivation is ``multimodal'' and ``language-conditioned''}

% what is texture authoring
Advances in haptic technologies and multimodal generation have increased the demand for methods to deliver realistic and congruent multisensory feedback, particularly centered around haptic signals, %in virtual environments 
through intuitive interfaces~\cite{theivendran2023rechap, wang2024genartist}. Previous work in virtual haptic textures has relied on the use of data recorded during interactions with real objects~\cite{romano2011creating}. While these methods can create realistic and multimodal interactions, they rely on collected texture samples and therefore cannot easily replicate new, unmodeled textures. Increasingly, users seek systems that use simple commands to seamlessly generate haptic cues that are aligned with other content, a process commonly referred to as \textit{texture authoring}, for use in applications like digital design, virtual prototyping, gaming, and medical simulation and training~\cite{basdogan2004haptics, xia2013review, choi2012vibrotactile}.

% mimic and reproducing real-world haptic sensations - exhaustive search as authoring
\oldtext{Early approaches to haptic texture authoring framed the process as one of recreation through modeling and rendering, where the objective was to reproduce the tactile sensations experienced when interacting with real-world textured surfaces~\cite{culbertson2014modeling}. In this paradigm, authoring was achieved by searching for a target texture within a library of pre-recorded physical samples that best matched the desired perceptual qualities. While effective for faithful reproduction, this method inherently constrained creative exploration, as it was limited to the tactile interactions and materials available in the physical environment.
Despite efforts to create diverse datasets and libraries~\cite{seifi2015vibviz, culbertson2014one}, this method remains \emph{data-bounded}: the haptic cues a system can deliver are limited by what has been recorded, curated, and parameterized.}
\newtext{Early authoring approaches focused on recreating real-world textures~\cite{culbertson2014modeling} by searching libraries of pre-recorded physical samples~\cite{seifi2015vibviz, culbertson2014one}. While effective for faithful reproduction, this \emph{data-bounded} paradigm inherently limits creative exploration to physically available materials, preventing the synthesis of novel or imaginary sensations.}

% move beyond mimicking; correlate texture properties with haptic signals
To move beyond direct mimicking and reproducing sensations from real-world textured materials, 
recent research has advanced toward correlating texture properties with %other 
sensory signals. This enables the creation of experiences that are difficult to capture directly or can only be inferred and grounded through other modalities (e.g., semantic meaning, vision, or language)~\cite{okamoto2012psychophysical, ujitoko2018vibrotactile, sung2025hapticgen}. 

%researchers have explored creating experiences that are difficult to record directly or can only be inferred and grounded from other modalities (\textit{e.g.,} semantic meaning, preference, language) by correlating texture properties with actual sensory signals~\cite{ , , }.  
Such approaches make it possible to synthesize new virtual sensations by manipulating the underlying texture properties, rather than relying solely on pre-recorded datasets. %in order to synthesize new virtual sensations from interactions with textures by manipulating the texture properties. 
However, key challenges arise at both ends of this correlation. On the one hand, texture properties are often perceived subjectively~\cite{okamoto2012psychophysical}. Although measurable descriptors such as roughness, compliance, or friction can be defined, users may interpret them differently depending on context and prior experience~\cite{tiest2010tactual}, making it difficult to ground texture properties with consistent perceptual or semantic mappings. Conversely, accurately capturing texture properties requires action-conditioned, multimodal, and crossmodally redundant sensory signals (e.g., tapping force to convey hardness, vibrations to represent roughness, and visual cues to suggest gloss or microstructure). Incorporating multiple modalities not only increases the richness of the signal representation but also introduces greater data complexity, more demanding collection procedures, and most critically, the challenge of aligning information across modalities.

% talk about prior work on texture authoring and their limitations; then transit to the multimodal and language-condition features of this work
% Facing the challenges

% linear interpolation, affective space, 

% preference-driven

Early efforts in this direction established the concept of texture authoring by linking the affective space derived from human perceptual ratings with the haptic model space, from which new textures can be rendered by interpolation~\cite{8710002}. 
With the rise of generative models, researchers began to leverage the traversals in latent spaces of these models to synthesize and tune haptic signal outputs, guided by adjective-based interpolation~\cite{8710002}, application needs~\cite{tozuka2025integrating}, or user preference~\cite{9772285}. 
Notably, to maintain the active exploration and interactivity of the haptic experience, several works used parametric haptic texture models as the generation target, rather than the raw haptic signals~\cite{8710002, 9772285}, emphasizing the importance of representations for real-time rendering and user-centered control and expressivity in the authoring process.  
However, while these approaches allow systems to generate haptic cues that extend beyond what has been physically measured, the grounding of generated textures (e.g., their correspondence to human perception and semantic meaning) has not been systematically established. Additionally, multimodal generation and alignment %of generated outputs 
across haptic and other modalities are still underexplored, with most existing work focusing on a single output modality related to textures~\cite{ujitoko2018vibrotactile,9197447,sung2025hapticgen}. These gaps limit the applicability and generalizability of free-form open-vocabulary texture authoring and constrain the potential for intuitive and crossmodal interactions with virtual textures.

To achieve this, we propose a \emph{multimodal variational autoencoder (VAE)} with a \emph{language-aligned shared latent space} that spans both haptic modalities above: \emph{tapping transients} and \emph{%action-conditioned 
sliding-produced vibrations}. %(AR coefficients and variance) 
A text encoder (CLIP-style Transformer) maps language prompts into this latent space and is trained via contrastive and conditional objectives to align linguistic semantics with action-conditioned haptic structure~\cite{Radford2021CLIP}. In parallel, a \emph{decoupled diffusion pipeline} generates \emph{texture images} conditioned on the same latent and text prompt, leveraging advances in text-to-image diffusion models. The shared, language-aligned latent thus serves as a hub for text-to-trimodal generation, cross-modal completion (e.g., encode any subset, decode the rest), and counterfactual composition (e.g., ``tap like X but slide like Y''). \newtext{While our system uses a stylus-based interface, this modality is critical for professional domains such as digital industrial design (e.g., prototyping the feel of consumer electronics), virtual surgery (where tool-tissue interaction is fundamental), and digital artistry (enhancing the tactile feedback of digital brushes and sculpting tools).}

The key contributions are as follows:

\begin{enumerate}
  \item \textbf{Language-guided, tri-modal generation:} Proposed a VAE that jointly encodes \emph{tapping} and \emph{sliding} when interacting with textured surfaces into a shared, \emph{language-aligned} latent space, enabling text-to-haptics (tap \& texture) generation and fine-tuned, diffusion-based text-to-image generation with cross-modal coherence.
  \item \textbf{Visual-haptic alignment via decoupled diffusion:} Conditioned image diffusion models on the shared, language-aligned latent space to generate texture images whose appearance is perceptually consistent with their rendered haptic responses. 
  \item \textbf{Evaluation of perceptual coherence and language-guided usability:} Conducted two-phase human-subject studies assessing (i) perceptual realism of VAE-generated textures through latent interpolations, and (ii) the usability test of language-based texture authoring in end-to-end interaction.
\end{enumerate}

%% file: sections/2-related.tex
\section{Related Work}
% Multimodal Texture Modeling and Rendering?
% \subsection{Texture Rendering (non-generative)}
\subsection{Multimodal Texture Rendering}
% texture images
% sliding-produced vibrations
% tapping transient
% frictions
% audio
Multimodal interaction with virtual environments has recently received increasing attention, particularly within the fields of haptics, virtual reality, and human–robot interaction, and continues to be an active topic of research~\cite{culbertson2018haptics, 11174044}. Vision shapes expectations, while touch verifies them through actions~\cite{gibson1962observations, heller1982visual, ERNST2004162, klatzky2010multisensory}. 
Visual feedback provides rich spatial and material cues such as gloss, roughness, and microstructure that strongly influence users' expectations of touch~\cite{xiao2016can}. In virtual textured surfaces, rendering techniques such as normal mapping~\cite{koniaris2014survey}, reflectance modeling~\cite{10.1145/300776.300778}, and photorealistic shading~\cite{10.1145/3272127.3275066} are employed to convey fine surface detail and material appearance. Beyond realism, recent work explores how visual cues can be parameterized or learned from multimodal datasets to maintain consistency with haptic feedback~\cite{FANG2024106088}.

For haptic feedback, the bidirectional nature in interaction determines that a variety of texture properties are best understood through active exploration, either through a tool or through bare fingers. For example, short impacts shape hardness impressions, while sliding excites frequency-dependent vibrations that carry fine roughness~\cite{okamoto2012psychophysical,manfredi2014natural}. %NOT DONE 

Extensive research has sought to reproduce such transient and vibrational sensations, which arise from impact and sliding, respectively, in virtual environments using standardized tools and setups, with approaches progressing from physics-based model to data-driven techniques~\cite{stefani2025signal, chen2025systematic}. Physics-based models create the signal outputs following the contact dynamics between the tool/finger and the target surfaces to be modeled~\cite{951362, 6548426}. Although these models build explicit relations between signal outputs with physical parameters such as surface geometry, material properties, and interaction forces, they often rely on simplified assumptions about contact mechanics %(e.g., linear elasticity) 
and require precise knowledge of boundary conditions. As a result, while they offer interpretability and control, their generalization and rendering precision to complex real-world textures and exploratory behaviors remain limited. 

Compared to physics-based methods, data-driven approaches directly map human actions to haptic signals in a black-box manner, relying on statistical models~\cite{6548424}, neural networks~\cite{7177703, 9662218, heravi2024development}, and time-series processes~\cite{Yi2023timeseriesdiffusion}. These methods greatly reduce the model tuning and enhance the rendering realism. However, their black-box nature %of data-driven mappings 
often limits interpretability and physical consistency, for instance, excessive parameter tuning can lead to model collapse and disrupt causal relationships between input actions and resulting sensory outcomes. 

A notable hybrid approach is the event-based force feedback for tapping proposed in~\cite{kuchenbecker2006improving}, which combines both physics-based and data-driven approaches: the tapping transients are modeled as decaying sinusoids inspired by physical observations, with parameters derived from recorded acceleration profiles. Similarly, for sliding-produced vibrations, \cite{romano2011creating} used autoregressive (AR) processes to capture high-frequency vibrations produced under specific force-speed conditions, with both force-speed selection and AR parameters computed from free-motion recordings of real textured surfaces. These works set the foundation for the haptic %texture
models adopted in this work, upon which we build multimodal texture generation from natural language with aligned visual outputs. 

% other modality candidate: macrostructure (pin array), friction, audio, thermal in one brief paragraph
While tapping transients and vibrations form the core of our approach, %we note that 
other modalities also contribute to texture perception. Frictional cues modulate the perceived stick–slip dynamics~\cite{4145211, 6548382, schwarz2016slip} while auditory feedback correlates with vibrations and impact to reinforce judgments of roughness and stiffness~\cite{8960441, 10972862}. Thermal signals convey information on thermal conductivity and surface finish quality~\cite{niijima2020thermalbitdisplay, lee2021three}. Proprioceptive and %feedback %about hand and finger position, together with 
cutaneous feedback further supports perception of macroscale surface features and overall object geometry~\cite{shen2023fluid, qian2024shape}. 
In this work, we focus on tapping transients and sliding-produced vibrations paired with visual previews as the basis for our multimodal texture generation and authoring.
% which directly contribute to our multimodal texture generation and authoring framework.

% \subsection{Generative and cross-modal models for tactile signals}
\subsection{Texture Generation and Authoring}
Texture generation and authoring initially drew inspiration from methods in image generation and authoring~\cite{xian2018texturegan, yang2023diffusion}. Over time, these methods have evolved from search-in-the-loop~\cite{koyama2020sequential}, which relied on user-guided retrieval and matching, to exploration within learned latent priors that enable free-form synthesis and semantic control~\cite{chen2023text2tex}.

% authoring by linear interpolation or sampling
Linearly interpolating the space derived either from affective descriptions or learned latent priors provides a straightforward way to create novel textures with intermediate perceptual qualities or semantics. \cite{8710002} created an authoring space by correlating affective attributes of physical textures with changes in haptic signals, allowing users to synthesize new textures that continuously vary along affective (perceptual) dimensions. \cite{10.1007/978-3-319-93399-3_3} fine-tuned the vibrotactile signals through linear sampling in the latent space of generative adversarial networks (GAN), showing that latent interpolation enables controlled manipulation of user perception on textures. However, the interpolation process in these methods is often unidimensional and requires subtle adjustments to reach the desired sensations, thereby hindering efficient texture authoring. 

% authoring by human preference or judgement
To effectively align the rendered haptic sensations with users' perceptual intent, preference-driven authoring leverages user's preference during the interactions with a set of texture candidates to navigate the GAN's latent space via an evolutionary strategy, converging toward the desired feel without tedious manual sampling or interpolation~\cite{9772285}.  
% uses a generator from a learned GAN model with evolutionary strategy %(CMA-ES) 
% to match a desired feel without tedious sampling or interpolation. %recordings~\cite{9772285}. 
To ensure interactivity during this process, this work uses the haptic texture models as the generation target, rather than the raw haptic signal. 
Subsequent research has built upon this idea by integrating human judgment or preference into the haptic authoring process, known as \textit{human-in-the-loop} framework, to produce outputs that better reflect subjective human perception~\cite{Zhang2024texasgan}

% authoring by other modality (vision, audio, xx) -> haptic cues
Design-oriented systems have increasingly linked visual and auditory modalities to expected tactile outcomes~\cite{faruqi2025tactstyle}. 
By leveraging cross-sensory correspondences and multimodal priors, these frameworks enable the generation of haptic feedback conditioned on perceptual cues across modalities, spanning generations of image$\rightarrow$friction or vibration~\cite{Cai2022image2friction,10160373,10496176}, 
audio$\rightarrow$vibration~\cite{Zhan2024audio2tactile}, 
% visual$\rightarrow$tactile variational autoencoding~\cite{Xi2024cmavae}, 
and video+audio$\rightarrow$friction%modeling% via Transformer architectures
~\cite{Song2023frictionTransformer}. 
% and vision-based haptic embeddings for robotic rendering~\cite{Cao2023vis2hap}.
Among these works, VAEs, GANs, and Transformer-based architectures dominate the design space, demonstrating strong cross-modal alignment.

% authoring via language 
Language has emerged as an effective authoring interface across 2D images~\cite{rombach2022latent,ramesh2022dalle2,saharia2022photorealistic}, 3D scenes~%with diffusion 
\cite{lin2023magic3d,xu2023dream3d}, and audio~\cite{kreuk2022audiogen,agostinelli2023musiclm,liu2024audioldm}.
Early efforts in haptics domain demonstrate the feasibility of text-to-haptic generation~\cite{tu2025texttoucher,sung2025hapticgen,11123227}, but remain limited in authoring %modality 
scope, typically focusing on a single haptic channel (often vibrations), and performing one-way translation rather than \emph{co-generation}.  
% We address these gaps by treating language as a \emph{pragmatic navigation} tool over an action-aware latent space. 
We address these gaps by learning a unified and physically grounded latent space that \emph{co-generates} both primary haptic channels, tapping transients and sliding-produced vibrations, paired with semantically aligned visual previews, all conditioned on natural language descriptions.

%% file: sections/3-method.tex
\section{Methodology}
\label{sec:method}

\begin{figure*}[t]
  \centering
  \includegraphics[width=1.0\linewidth]{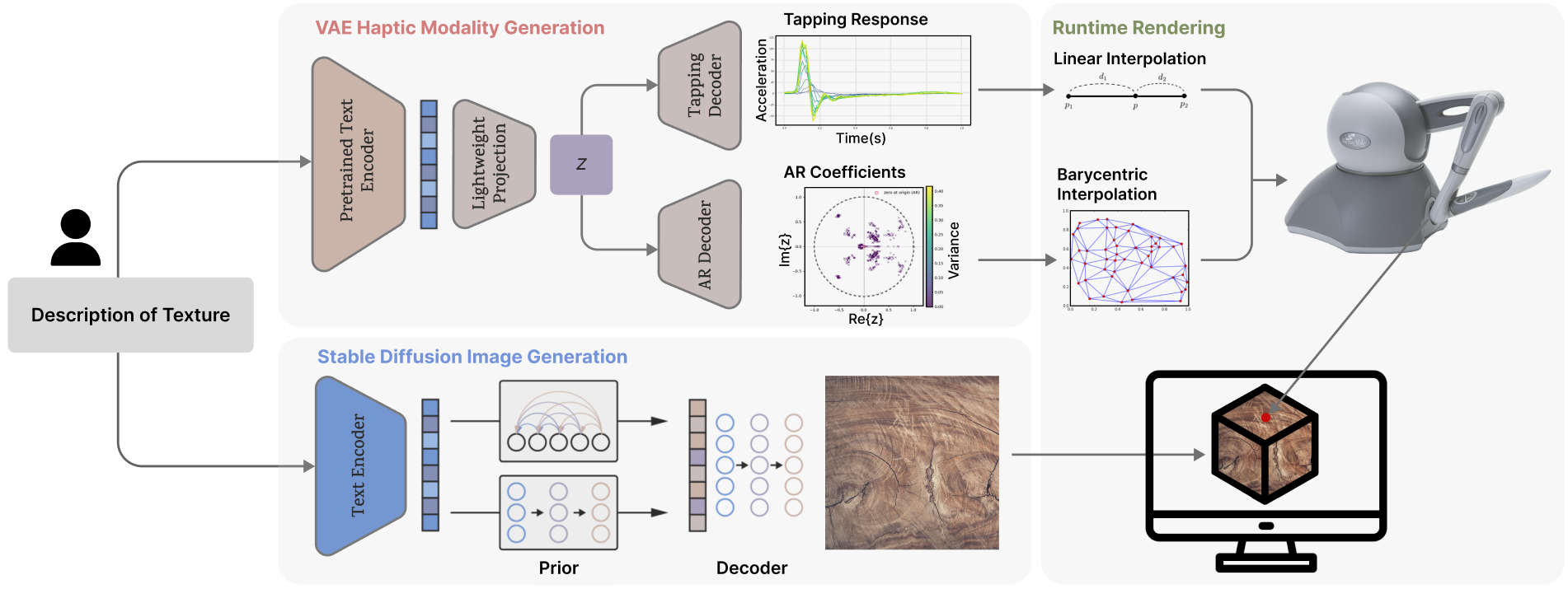}
  \caption{\textbf{System overview.} A text prompt is encoded into a haptic latent $z$. Modality-specific decoders then output (i) a tap bank for hardness rendering and (ii) an AR matrix for sliding-produced vibrations. The same prompt also drives a decoupled text-to-image diffusion model for the visual preview.}
  \label{fig:inference}
\end{figure*}

Our system turns a short text description of a texture into two haptic signals -- (i) a set of force/speed–conditioned autoregressive (AR) models to provide virtual texture vibrations during \emph{sliding} and (ii) a bank of event-based \emph{tapping} transients to provide hardness information -- and, in parallel, an \emph{independent} text-to-image generation for visual contexts. We first describe the end-to-end inference path (Fig.~\ref{fig:inference}), then how the latent is learned (Fig.~\ref{fig:training}), followed by data, objectives, and runtime rendering.

\subsection{Haptic Model Components}
\label{sec:haptic-model}
\paragraph{Sliding-produced vibrations}
Following the data-driven texture modeling framework proposed in~\cite{culbertson2014texture,culbertson2014modeling}, we represent each material using a set of stable, low-order autoregressive (AR) processes indexed by %instantaneous 
normal forces $f$ and tangential speeds $v$, which constitute a 2D $f{-}v$ grid. %(we use $M{=}18$ force-speed pairs). 
These force-speed pairs are segmented, and their corresponding AR parameters are computed from data collected during free-motion tool-surface interactions. 
%For numerical stability, we store each AR process in the form of \emph{line spectral frequencies} (LSFs) with one scalar \emph{excitation variance}. 
At runtime, we interpolate the AR coefficients according to the current force and speed within this grid and run the following inference procedure.
%compute the interpolated AR processes based on the location of the current force and speed in the grid and run the following inference procedure.
% and convert the LSFs back to AR coefficients.
\begin{equation}
\label{eq:ar}
\begin{split}
y[n] &= \sum_{k=1}^{p} a_k(f,v)\,y[n-k] + \varepsilon[n],\\
\varepsilon[n] &\sim \mathcal{N}\!\big(0,\sigma^2(f,v)\big)
\end{split}
\end{equation}
Here, $y[n]$ is the vibration signal at $n^{\text{th}}$ time step and $a_k(f,v)$ are the AR coefficients corresponding to the current force $f$ and speed $v$. %determines how vibration feels like, 
$\varepsilon[n]$ is a zero-mean random excitation, %—one new sample at each time step—
%which is updated at each time step 
of which variance $\sigma^2(f,v)$ controls the overall energy. Intuitively, $a_k(\cdot)$ sets the “timbre” (e.g., smoother vs.\ scratchier) and $\sigma^2(\cdot)$ sets the “loudness” (how strong the vibration feels). Thus, holding the AR coefficients $a_k(f,v)$ fixed, increasing $\sigma^2$ makes the texture feel more energetic or “bursty” (as with gritty abrasives), whereas decreasing it yields a quieter glide (as with polished or compliant surfaces). During inference, we continuously interpolate both $a_k$ and $\sigma^2$ from the recorded grid to match the user’s current $(f,v)$, so the produced vibrations using AR processes adapt to the user’s actions~\cite{culbertson2014modeling,culbertson2016importance}.

\paragraph{Tapping transients}
Hardness is conveyed by short, high-frequency transient forces at impact~\cite{culbertson2016importance,kuchenbecker2006improving}. The spectral centroid of these transients is roughly constant for a specific material, with high-frequency transients corresponding to stiff materials and vice versa for soft ones. %lower frequency transients corresponding to softer materials. 
The amplitude of the transients scales linearly with impact speed. For each material, we store $13$ recorded acceleration traces covering low to high impact speeds and linearly interpolate between them at runtime to match the measured impact speed. 

Sliding %(AR) 
and tapping %(transients) 
are rendered together so that users perceive both surface micro-geometry and substrate firmness. 
We discuss the force rendering of these components, including surface stiffness, in Sec.~\ref{sec:runtime}.

\subsection{Data, Representation, and Normalization}
\label{sec:data}
\paragraph{Corpus}
We use haptic models from the Penn Haptic Texture Toolkit (HaTT)~\cite{culbertson2014texture}: for each of 100 materials, we load (i) an XML with AR processes (stored as \emph{line spectral frequencies} (LSFs)) over a unified force–speed conditions 
($\text{number of conditions}=18$), and (ii) a tapping file with acceleration traces at $13$ different impact speed (each $T{=}100$ samples at $10$\,kHz) collected per~\cite{culbertson2016importance}. We note that the textures in HaTT are \textbf{isotropic} and \textbf{rigid}: (A) \emph{isotropy}, meaning that the rendered high-frequency %micro-
vibrations do not depend on in-plane sliding direction; (B) \emph{rigidity}, meaning that the substrate does not undergo large-scale deformation and the feel is dominated by surface asperities and contact dynamics. These assumptions allow us to focus on high-frequency vibration physics rather than large-scale geometric or structural effects. 

% match the HaTT corpus and keep the focus on micro-vibration physics only rather than including macro structure modeling.

\paragraph{Augmentation}
\oldtext{Because the HaTT includes a small number of models for each texture, capturing only the behaviors at limited interaction conditions, we must augment the models.}
\newtext{Because the HaTT corpus contains only 100 distinct materials, the resulting feature space is naturally sparse. To enable the model to learn a continuous manifold capable of smooth interpolation, we must augment the models to densify the training distribution.} To increase variability of the haptic models while preserving identity of the underlying materials, we implement (a) \emph{tap mixing}: linearly mix each tap set with traces from 19 distinct classes using high target weights (0.95:0.05), so small cross-material perturbations are added without erasing hardness cues; (b) \emph{AR resampling:} Following~\cite{9772285}, we first cluster all recorded force–speed samples in HaTT into 18 bins to define a unified $f{-}v$ grid across textures. For each texture, the AR model is augmented 20 times by sampling one entry from each bin to construct a new instance. Each sampled entry is labeled with the bin's centroid $f{-}v$ and its parameters are computed via interpolation. 
% we augment the AR models in HaTT by clustering all recorded force–speed samples into 18 bins and repeatedly resampling one entry per cluster to construct a new AR model. 
% Each sampled entry is labeled with its cluster centroid and computed via interpolation. %used to interpolate new AR models. 
This yields $2{,}000$ (AR, tap) training pairs.

 \paragraph{Tensors and normalization}
Each item is stored as an \textbf{AR tensor} of size $18{\times}(21{+}1)$ (21 LSFs $+$ one variance for each of the 18 force-speed conditions) and a \textbf{tap tensor} of size $13{\times}100$ (100-datapoint stream for each of the 13 impact speeds). We apply channel-wise $z$-normalization which keeps each channel on a comparable numerical scale across all materials, preventing high-magnitude features from dominating learning.
\oldtext{All $2{,}000$ pairs are used for training; the lack of zero-shot held-out materials is discussed in Sec.~\ref{sec:discussion}.}
\newtext{We utilize the full dataset of 2,000 pairs for training without a held-out zero-shot set. Given the limited class count, withholding distinct textures would create large semantic gaps in the latent space, severely impairing the model's ability to extrapolate or interpolate between material categories.}

\subsection{System Architecture}
\label{sec:inference}
Fig.~\ref{fig:inference} showcases the runtime pipeline. A user provides a short description (e.g., ``rough abrasive sheet with gritty particles''), which is encoded into a feature vector by a frozen \texttt{CLIP ViT-B/32} encoder. A lightweight projection head transforms this vector into a haptic latent $z\in\mathbb{R}^{64}$. Two decoders then generate: 
\begin{itemize}
\item an \textbf{AR matrix} of size $18{\times}(21{+}1)$ (LSFs $+$ variance at each force–speed condition), and
\item a \textbf{tap bank} of size $13{\times}100$ (100 post-contact samples at 13 impact speeds).
\end{itemize}
These outputs are streamed to the haptic renderer to produce kinesthetic and vibrotactile feedback via a customized haptic device. In parallel, the same text drives a diffusion model to render a visual preview. The visual and haptic branches are \emph{decoupled} at inference; cross-modal coherence arises from the \emph{shared language prompt}, as discussed in the following subsection.

\begin{figure}[t]
  \centering
  \includegraphics[width=1.0\linewidth]{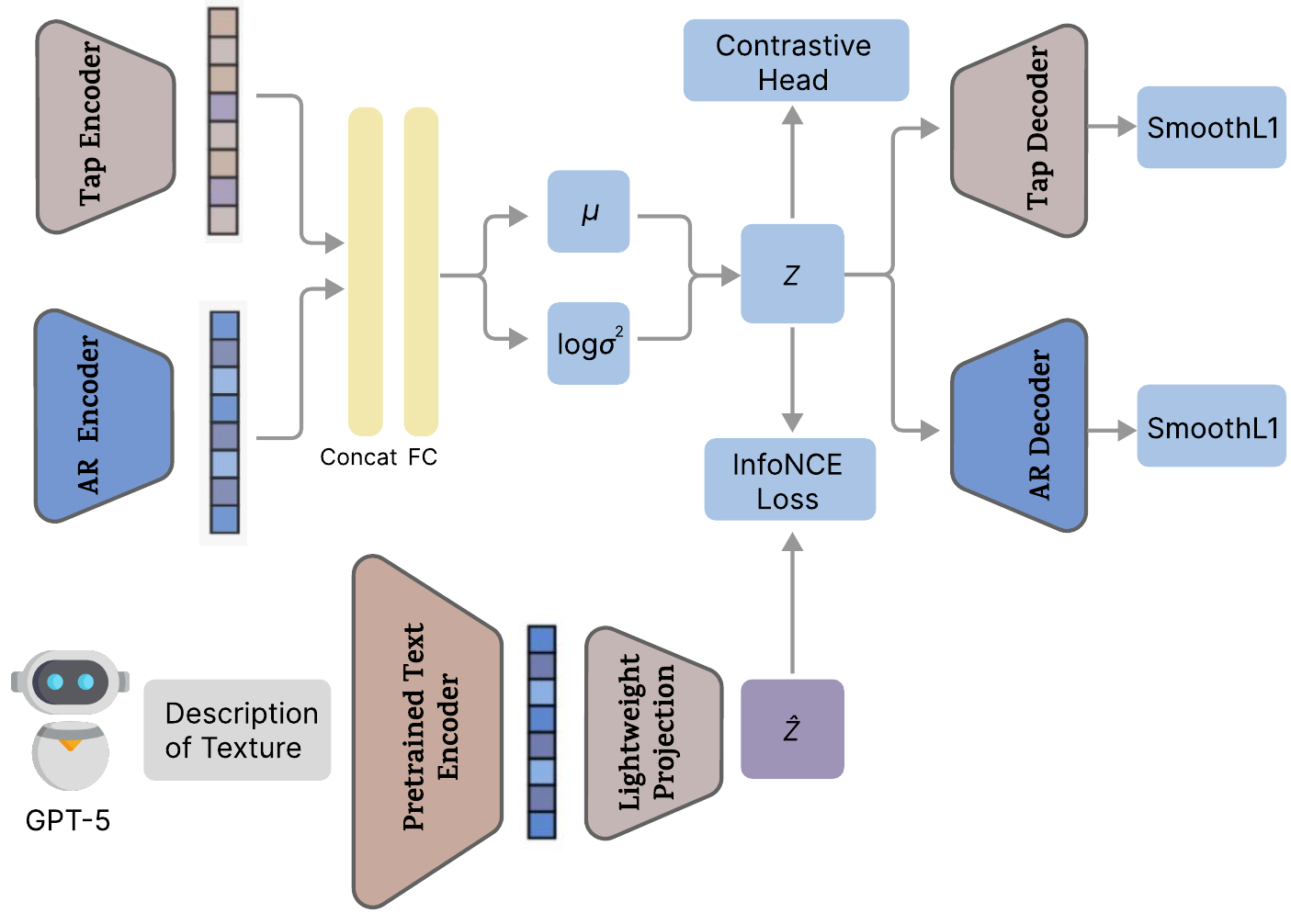}
  \caption{\textbf{Learning latent.} A bimodal VAE reconstructs tap bank and AR matrix from haptic latent $z$; a contrastive head aligns $z$ with CLIP text, with KL regularization keeping the latent compact and reconstruction losses preserving haptic fidelity.}
  \label{fig:training}
\end{figure}

\subsection{Learning Haptic Latent}
\label{sec:training}
% \paragraph{Goal}
We aim to learn a unified, \emph{action-aware} latent that co-generates tapping transients and sliding vibrations, and can be \emph{steered by language}. We describe inputs, network, and objectives as follows (Fig.~\ref{fig:training}).

\paragraph{Inputs and text supervision}
% Each training sample provides an AR tensor and a tap tensor (Sec.~\ref{sec:data}). For language, five short captions per material generated by GPT5~\cite{openai_chatgpt_2025}. Prompt for generating captions include texture images from HaTT dataset combined with examples including tactile adjectives (e.g., ``rough, gritty sandpaper texture; crinkled metallic foil surface that feels sharp and thin''); one caption is sampled per step. Captions are embedded by a frozen CLIP text encoder~\cite{Radford2021CLIP}.

Each training sample offers an AR tensor and a tap tensor (Sec.~\ref{sec:data}).  
For language conditioning, we generate five short captions per material using \texttt{GPT-5}~\cite{openai_chatgpt_2025}, guided by \textit{visual image grounding} and \textit{few-shot expert prompting}. Specifically, the prompt for generating caption uses a template that includes all of the following:  
(i) the corresponding texture image from the HaTT as visual context,  
(ii) three expert-crafted examples containing tactile adjectives and multimodal phrasing  
(iii) explicit guidance to vary linguistic focus across trials—covering visual appearance, tactile sensation, and material composition—to encourage descriptive diversity.  

This setup ensures that generated captions are both semantically rich and perceptually relevant, capturing haptic qualities grounded in visual texture appearance rather than generic language priors.  
During training, one caption is sampled per step, and all captions are embedded using a frozen \texttt{CLIP ViT-B/32} text encoder~\cite{Radford2021CLIP}.

\paragraph{Network}
As shown in Fig.~\ref{fig:training}, a \emph{bimodal VAE} encodes taps and AR tensors into a shared Gaussian posterior $q_\theta(z|x)$ over $z\in\mathbb{R}^{64}$; modality-specific decoders map $z$ back to a tap bank and an AR matrix. % (LSFs $+$ variance). 
Two small projection heads (one from $z$, one from $\hat z$) are L2-normalized so that a contrastive loss can align matched (text, $z$) pairs. Details are listed in Table~\ref{tab:spec_summary}. 

\paragraph{Objectives}
Our training optimizes three terms: (i) \textbf{Reconstruction}: a Smooth-$L_1$ loss on taps and AR tensors to ensure decoded signals match the dataset; (ii) \textbf{Latent regularization}: a KL term that keeps $z$ compact and well-behaved; and (iii) \textbf{Language alignment}: a InfoNCE with in-batch negatives to align text embeddings with the corresponding material latents. Concretely, with reconstructions $\hat x_{\text{tap}},\hat x_{\text{AR}}$, and L2-normalized projections $e_x$ (latent) and $e_t$ (text), we minimize
\begin{align}
\mathcal{L} &= \lambda_{\text{rec}}\!\left(\lambda_{\text{tap}}\|x_{\text{tap}}-\hat x_{\text{tap}}\|_1
+\lambda_{\text{ar}}\|x_{\text{AR}}-\hat x_{\text{AR}}\|_1\right) \notag\\
&\quad + \beta(t)\,D_{\text{KL}}\!\bigl[q_\theta(z|x)\,\|\,\mathcal{N}(0,I)\bigr] \notag\\
&\quad + \lambda_{\text{text}}\,\mathcal{L}_{\text{InfoNCE}}(e_x,e_t;\tau)
+ \lambda_{\text{align}}\|\hat\mu-\mu\|_2^2,
\label{eq:loss}
\end{align}

\noindent with $\tau{=}0.1$, batch size $32$, $\lambda_{\text{tap}}{=}\lambda_{\text{ar}}{=}2.0$, $\lambda_{\text{text}}{=}0.1$, and $\beta(t)$ linearly annealed from $0$ to $0.001$ over 20–120 epochs~\cite{higgins2017beta}. We use Adam optimizer (learning rate $10^{-4}$) with gradient clipping at $1.0$. 

% \paragraph{Specification summary (for reproducibility)}
% \vspace{2mm} % or \smallskip
% \noindent

\begin{table}[t]
\centering
\caption{Specification Summary}
\renewcommand{\arraystretch}{1.2}
\begin{tabular}{@{} >{\bfseries}r @{\hspace{0.5em}} p{0.82\linewidth} @{}}
\toprule
Latent: &
$d_z{=}64$ Gaussian $(\mu,\log\sigma^2)$. \\

AR encoder: &
$\mathbb{R}^{18\times 22}\!\rightarrow\!396\!\rightarrow\!256\!\rightarrow\!256\!\rightarrow\!128\!\rightarrow\!64$ (ReLU). \\

Tap encoder: &
$\mathbb{R}^{13\times 100}\!\rightarrow\!1300\!\rightarrow\!256\!\rightarrow\!64$ (ReLU). \\

Decoders: &
AR: $64\!\rightarrow\!256\!\rightarrow\!512\!\rightarrow$[3$\times$Res(512, ReLU, Dropout 0.1)] with two heads (LSF $18{\times}21$ and variance $18{\times}1$). 
Tap: $64\!\rightarrow\!256\!\rightarrow\!512\!\rightarrow$[2$\times$Res(512)] $\rightarrow 1300$. \\

Text side: &
CLIP ViT-B/32 (frozen) with MLP proj: $512\!\rightarrow\!256\!\rightarrow\!64$; latent proj: $64\!\rightarrow\!128\!\rightarrow\!64$ (BN, Dropout 0.1), both L2-normalized. \\

Note: &
To guarantee stability, we output LSFs via softmax$\!\to$cum-sum, scale to $(0,\pi)$, and clamp to $\pi{-}10^{-4}$ before LSF$\!\to$AR conversion. \\
\bottomrule
\end{tabular}
\label{tab:spec_summary}
\end{table}

\subsection{Image Generation (decoupled visual preview)}
\label{sec:image}
For the preview image, we use an off-the-shelf \texttt{Stable Diffusion 2}-base pipeline~\cite{stabilityai_stableDiffusion2Base_huggingface}, optionally fine-tuned with PolyHaven textures \cite{polyhaven_textures}. The image is not conditioned on the haptic latent $z$; cross-modal coherence emerges because both haptic modalities and image are driven by the same language description.

\subsection{Runtime Haptic Rendering}
\label{sec:runtime}

We deploy on a 3D Systems Touch device at a $1$\,kHz servo loop. At each tick, the device state (pose, penetration depth $\delta$, tangential velocity $\mathbf{v}_t$ with speed $v_t=\|\mathbf{v}_t\|$) is measured and used to compute the output force $\mathbf{F}$. The total force applied at the stylus is
\begin{equation}
\label{eq:totalforce}
\mathbf{F} \;=\; F_n\,\hat{\mathbf{n}}
\;-\;F_t\,\hat{\mathbf{t}}
\;+\;F_\text{vib}\,\hat{\mathbf{t}}
\end{equation}
where $\hat{\mathbf{n}}$ and $\hat{\mathbf{t}}$ are the surface normal and a unit tangential direction of the velocity at the contact point, respectively. $F_n$ is normal contact force, $F_t$ is friction force, and $F_\text{vib}$ is vibration-induced force. We now define each term.

\subsubsection{Normal contact (surface stiffness)}
The normal term models surface stiffness via a virtual spring:
\begin{equation}
\label{eq:normal-force}
F_n \;=\; k_n\,\delta
\end{equation}
with virtual stiffness $k_n>0$ and penetration depth $\delta \ge 0$. During impacts, a transient normal impulse is added (Sec.~\ref{sec:runtime}-\textit{Tapping transients}).

\subsubsection{Sliding vibrations (texture component)}
When $\delta>0$, the decoded AR matrix provides force/speed–conditioned contact dynamics. We barycentrically interpolate the AR coefficients and excitation variance at the current contact condition $(F_n,v_t)$:
\[
\{a_k(F_n,v_t),\,\sigma^2(F_n,v_t)\}
\]
which is used to synthesize the vibration signal $y[n]$ by the AR recursion in Eq.~\ref{eq:ar}. 
% The vibration sample is synthesized by the AR recursion in Eq.~\ref{eq:ar}. 
This signal is scaled by the device gain $g_\text{vib}$ and the maximum continuous output $F_{\max}$ to output vibration-induced force $F_\text{vib}$:
\begin{equation}
\label{eq:vibforce}
F_\text{vib} \;=\; g_\text{vib}\,y[n]\,F_{\max}
\end{equation}

\subsubsection{Friction rendering (tangential damping)}
To render slipperiness, we estimate a friction coefficient from nearby materials in the latent space. With an anchor set 
$\mathcal{A}=\{(\mathbf{z}_i,\mu_i)\}$ of paired latent $z_i$ and friction coefficient $\mu_i$, the coefficient for a query latent $\mathbf{z}_q$ is
\begin{equation}
\label{eq:mu-est}
\mu(\mathbf{z}_q)=
\frac{\sum_{i\in\text{top-}k}\exp(\cos(\mathbf{z}_q,\mathbf{z}_i)/\tau)\,\mu_i}
     {\sum_{i\in\text{top-}k}\exp(\cos(\mathbf{z}_q,\mathbf{z}_i)/\tau)} ,
\end{equation}
where $\tau>0$ controls weighting softness. The friction force in Eq.~\ref{eq:totalforce} is then
\begin{equation}
\label{eq:friction}
F_t \;=\; \,\mu(\mathbf{z}_q) F_n
\end{equation}

\subsubsection{Tapping transients (hardness cue)}
For impacts, each texture includes 13 precomputed acceleration traces across contact speeds. The impact speed $v_\text{tap}$ is estimated from normal motion; the transient acceleration $a_\text{tap}(t)$ is obtained by linear interpolation between the two nearest traces in $v_\text{tap}$. The corresponding normal impulse for effective device mass $m_\text{eff}$ is
\begin{equation}
\label{eq:tapforce}
F_\text{tap}(t) \;=\; m_\text{eff}\,a_\text{tap}(t)
\end{equation}
which is added to the normal force during the impact window:
\begin{equation}
    F_n \;\leftarrow\; F_n + F_\text{tap}(t)
\end{equation}

% \]

%% file: sections/4-result.tex
\section{Experimental Results}
We assess whether the learned haptic latent (trained on AR matrix~+~tap bank only) captures perceptually consistent relationships among materials; whether text prompts generate plausible tri-modal  outputs (two haptic channels, and one visual); and whether interpolations within the latent space produce smooth and interpretable transitions aligned with human perception.

\subsection{Haptic Latent Assessment}
\label{sec:tsne}

\begin{figure}[t]
    \centering
    \includegraphics[width=\linewidth]{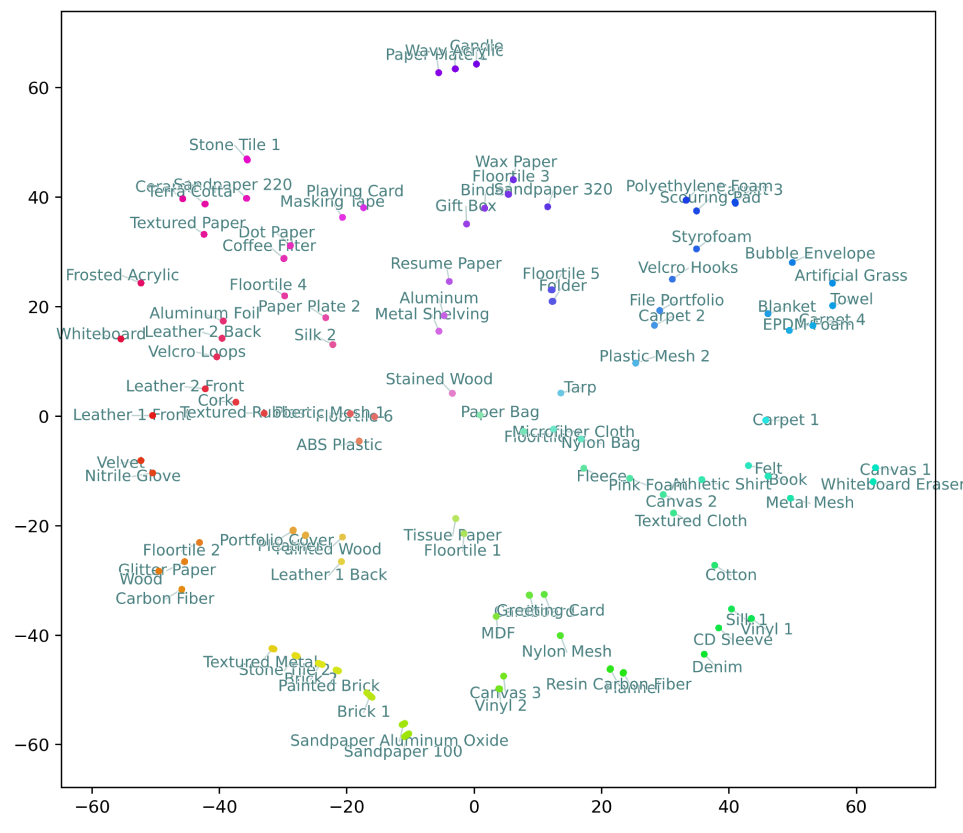}
    \caption{\textbf{VAE latent space embedding.} Each point represents the posterior mean $\boldsymbol{\mu}$ for one of the 100 texture classes after dimensionality reduction (PCA + t-SNE). Nearby points correspond to materials that the model internally considers perceptually similar.}
    \label{fig:latentspace}
\end{figure} 

\subsubsection{Qualitative visualization}
For each of the 100 textures in HaTT, we extract the posterior mean vector $\boldsymbol{\mu}$ from the learned latent representation, which compactly encodes each material’s vibration and impact dynamics. To visualize these high-dimensional features, we first apply Principal Component Analysis (PCA) to remove minor noise and emphasize dominant variance directions, followed by t-distributed Stochastic Neighbor Embedding (t-SNE) for two-dimensional projection.  
t-SNE preserves local similarity, such that materials positioned close together share similar internal representations, while distant points indicate strong perceptual differences.
% t-SNE preserves local similarity: materials mapped close together share similar internal representations, while distant points indicate strong perceptual differences.

\begin{figure*}[tb]
    \centering
    \includegraphics[width=0.9\linewidth]{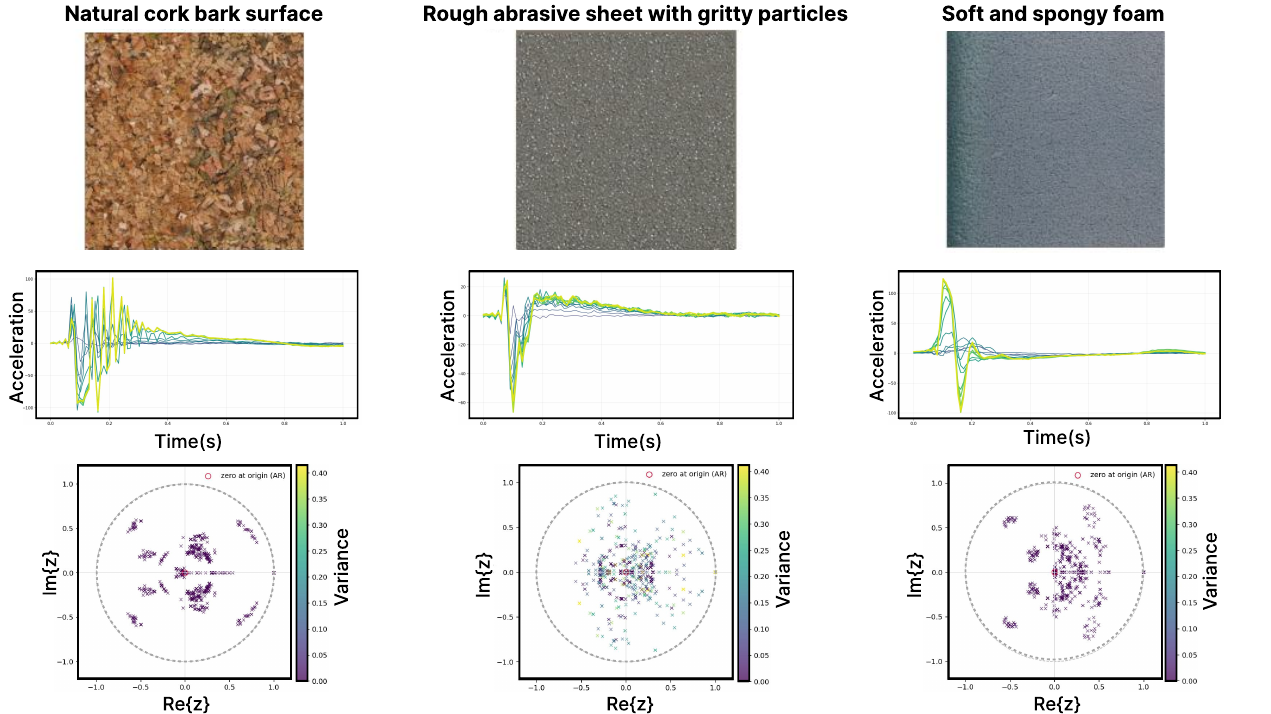}
    \caption{Tri-modal texture generation from language.
    \textbf{Top}: image generated by a diffusion model.
    \textbf{Middle}: 13 synthesized tapping responses, where color saturation indicates higher impact velocity.
    \textbf{Bottom}: pole distributions of the generated AR models plotted on the unit circle, with marker color denoting excitation variance.}
    \label{fig:exampleoutput}
\end{figure*}

Fig.~\ref{fig:latentspace} reveals clear structure in the model’s internal representation. For instance, soft foams cluster near other compliant materials, abrasive papers group tightly together, and metals or hard plastics occupy a distinct region—showing that the model organizes textures according to underlying perceptual attributes and physical properties.  
In other words, proximity in latent space corresponds to similarity in how materials feel through touch.

% \textbf{Quantitative validation.}
\subsubsection{Quantitative validation} 
To confirm that the observed patterns reflect genuine structure rather than visual artifacts, we evaluate cluster compactness and alignment using standard metrics from representation learning~\cite{mukherjee2019clustergan,xie2016unsupervised,ben2018gaussian}:

\begin{itemize}
  \item \textbf{Cluster compactness.}  
  High internal consistency (\emph{Silhouette} $\approx 0.96$, \emph{Calinski–Harabasz} $\approx 2.3{\times}10^4$) indicates that materials grouped together in latent space also share highly similar haptic behaviors.
  \item \textbf{Category separation.}  
  A low \emph{Davies–Bouldin} index ($\approx 0.14$) confirms minimal overlap between clusters. Materials that feel different remain well separated in representation.
  \item \textbf{Agreement with ground truth.}  
  The model’s clusters align strongly with the 100 labeled texture classes (\emph{Adjusted Rand Index} $\approx 0.97$, \emph{Normalized Mutual Information} $\approx 0.99$), implying that the latent organization closely parallels human-defined categories.
\end{itemize}
 
This structural organization underpins subsequent interpolations and user evaluations. The smooth, semantically ordered latent geometry enables coherent generation paths (Sec.~\ref{sec:interp}) and supports human evaluation (Sec.~\ref{sec:phase1}) to probe how users perceive transitions between nearby regions in this space.

\begin{figure*}[t]
    \centering
    \includegraphics[width=0.9\linewidth]{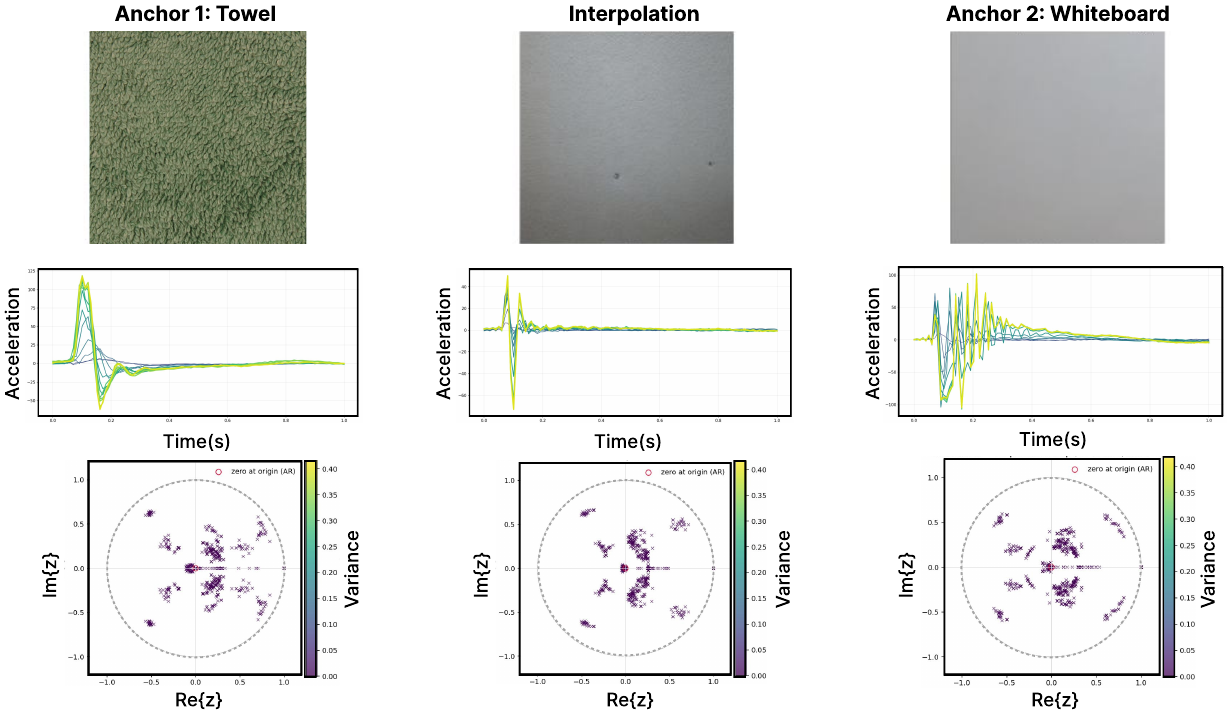}
    \caption{\textbf{Latent interpolation exemplar.} Columns show \textsc{Towel}$\rightarrow$latent midpoint$\rightarrow$\textsc{Whiteboard}. 
    \emph{Top}: diffusion image (semantics only, no imposed orientation). 
    \emph{Middle}: synthesized tap responses across 13 impact speeds (higher saturation = higher speed). 
    \emph{Bottom}: AR pole distributions on the unit circle, colored by excitation variance.}
    \label{fig:interpolres}
\end{figure*}

\subsection{End-to-End Generation}
\label{sec:e2e}

To validate the full text-to-tri-modal pipeline, we show three representative results generated end-to-end from prompts (\emph{“natural cork bark surface,” “rough abrasive sheet coated with gritty particles,”} and \emph{“soft and spongy foam”}) by visualizing the Images, Tapping Transient, AR models in Fig.~\ref{fig:exampleoutput}. The generated images are semantically aligned and directionally neutral, consistent with the isotropy assumption in rendering. The haptic signals exhibit material-appropriate signatures:
\begin{itemize}
\item \textbf{Cork:} Taps show broader, quieter onsets with rapid damping across all speeds. AR poles cluster further from the unit circle with moderate variance, consistent with the compliant, cellular structure of the material.
\item \textbf{Abrasive sheet:} Taps produce the sharpest onsets and shortest decays, particularly at higher speeds. AR poles spread toward the unit circle with elevated variance, reflecting collision-dominated dynamics at micro-asperity contact points.
\item \textbf{Foam:} Taps display the softest, most spread-out onsets with gradual decays characteristics. AR poles concentrate deep within the unit circle with minimal variance, consistent with the material's strong energy-absorbing properties.
\end{itemize}

Across prompts, these patterns yield the expected ordering in spectral sharpness and excitation strength (Abrasive sheet $>$ Cork $>$ Foam) while preserving within-sample speed progressions in the tap stack. The variance coloring in the bottom row distinguishes materials dominated by stochastic micro-collisions (Abrasive sheet) from those exhibiting more damped responses (Cork, Foam).

% The bottom-row variance coloring highlights where excitation is governed by stochastic micro-collisions (Abrasive sheet) versus damped responses (Cork, Foam).

\subsection{Latent Space Interpolation}
\label{sec:interp}

We visualize how the decoded signals evolve when interpolating around the average latent between two real textures. 
Given the posterior means $\boldsymbol{\mu}_A$ and $\boldsymbol{\mu}_B$ for two anchors, we compute their average
\begin{equation}
\bar{\mathbf{z}} = \frac{1}{2}(\boldsymbol{\mu}_A + \boldsymbol{\mu}_B),
\label{eq:avg_latent}
\end{equation}
and sample nearby latent around $\bar{\mathbf{z}}$ to generate the corresponding haptic reconstructions from the VAE decoder.
Visuals are generated by diffusion using the corresponding text prompt (e.g., “texture between \textit{towel} and \textit{whiteboard}”). 

Fig.~\ref{fig:interpolres} presents \textsc{Towel}$\rightarrow$\textsc{Whiteboard} as a representative path; we observe similar behaviors on other pairs. 
We note that the user study in Sec.~\ref{sec:phase1} analyzes how participants perceptually judge such generated latent interpolations; here we only visualize the signals produced by the model.

From towel to the latent midpoint, the AR spectra evolve in a structured but non-affine manner. Low-band energy associated with fibrous drag gradually recedes, while high-band content linked to polished sliding becomes increasingly prominent toward the whiteboard end. This transition follows a curved trajectory rather than a straight interpolation: mid-frequency regions adjust earlier than the highest bands, consistent with human roughness sensitivity peaking at mid frequencies and with regime mixing in the decoder, where the latent midpoint does not simply average AR coefficients across all force–speed settings.

Tapping transients at the midpoint exhibit a crisper onset and slightly longer ringing than towel, but are far less peaky than whiteboard, reflecting increased contact stiffness with residual compliance.

Visually, the diffusion-generated midpoint softens the towel weave and reduces the whiteboard sheen while maintaining isotropy. This demonstrates that the model preserves coherent, semantically aligned structure across modalities.

Across latent paths, both AR spectra and tapping transients move in material-plausible ways, while the corresponding images remain consistent with the underlying language semantics. However, these observations are signal-level; in the next section (Sec.~\ref{sec:phase1}), we present users’ perceptual ratings for these latent interpolations.

%% file: sections/5-usrEval.tex
\section{User Evaluation}
\label{sec:user-eval}

We conducted a two-phase user study ($N{=}$17; 10 male, 6 female, 1 non-binary). Twelve participants reported no prior experience with haptic devices (e.g., 3D Systems Touch, Novint Falcon). The study was approved by the University of Southern California Institutional Review Board under protocol UP-20-01131; all participants gave informed consent and participated voluntarily.

Our goals were to: (i) test whether the learned latent supports \emph{directional, semantically interpretable} changes between reconstructed anchors, and (ii) evaluate whether latent interpolations produce \emph{coherent, novel} textures that inherit recognizable properties from both anchors without collapsing to trivial averages.

\subsection{Anchor-Referenced Perceptual Interpolation}\label{sec:phase1}

\begin{figure}[tb]
    \centering
    \includegraphics[width=1\linewidth]{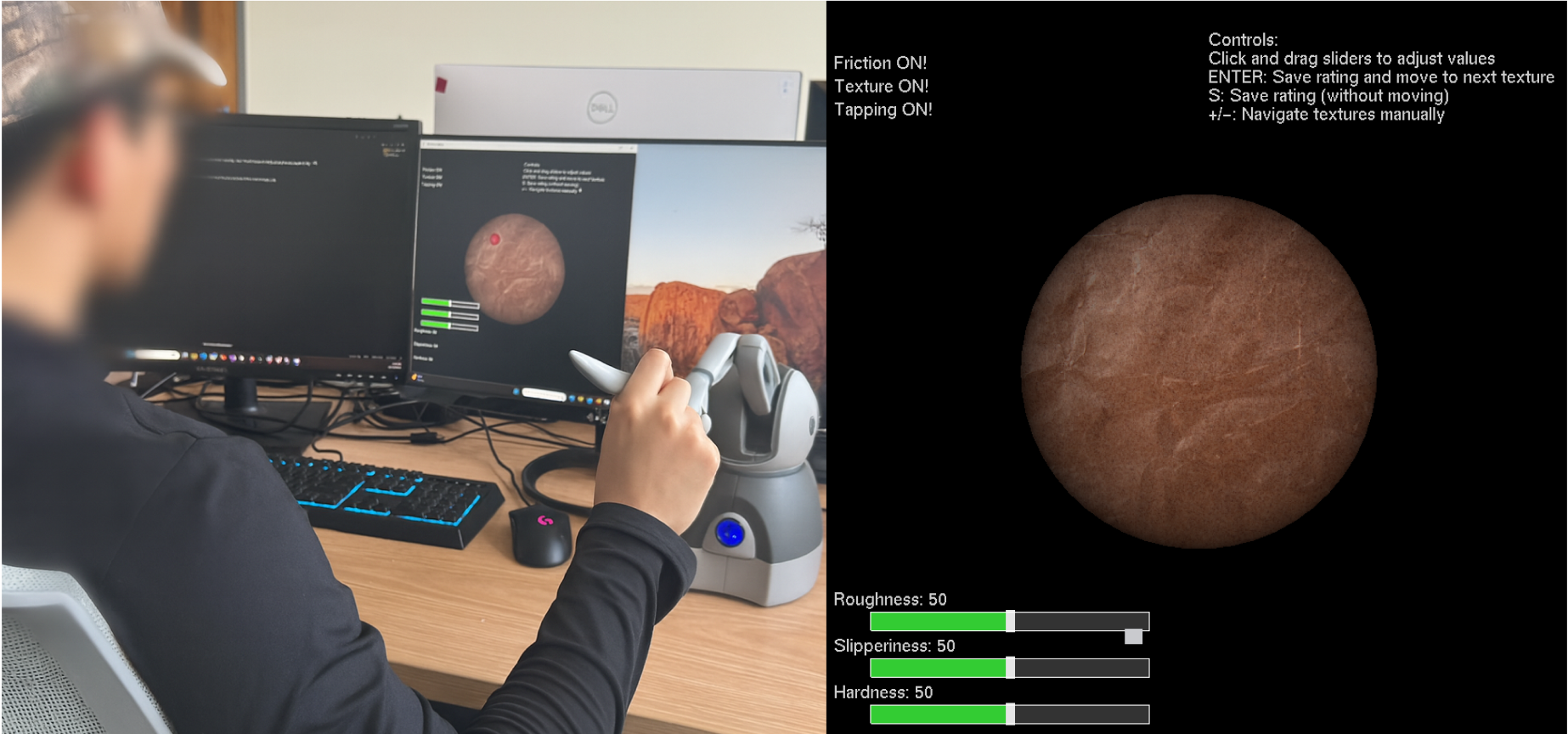}
    \caption{\textbf{User study setup.} \emph{Left:} Participant interacting with the 3D Systems Touch device while viewing the texture and rating UI. \emph{Right:} On-screen interface showing the rendered texture, controls, and three sliders for rating Roughness, Slipperiness, and Hardness.}
    \label{fig:user-setup}
\end{figure}

\textbf{Setup.} We formed all $\binom{5}{2}{=}10$ anchor pairs $(A,B)$ from \emph{Sandpaper}, \emph{Velvet}, \emph{Styrofoam}, \emph{Aluminum Foil}, and \emph{Velcro Hooks} (selected from HaTT Corpus for distinct perceptual features). For each pair, we generated the haptic intermediate by decoding the arithmetic mean of the two anchors’ encoder posterior means, and an image produced by diffusion model prompted as “texture between \emph{anchor1} and \emph{anchor2}.” as described in Sec.~\ref{sec:interp}. Participants explored each sample by sliding and tapping with a 3D Systems Touch device while viewing the corresponding texture image. They were asked to use three 0–100 sliders for the ratings of \textbf{R}oughness, \textbf{S}lipperiness, and \textbf{H}ardness for the given texture (Fig.~\ref{fig:user-setup}). \newtext{These dimensions were selected as they represent the principal axes of tactile perception of textures through a tool~\cite{okamoto2012psychophysical}, whereas attributes like temperature or macro-geometry were excluded due to hardware and modeling constraints. }

\begin{figure} [tb]
  \centering
  \includegraphics[width=\linewidth]{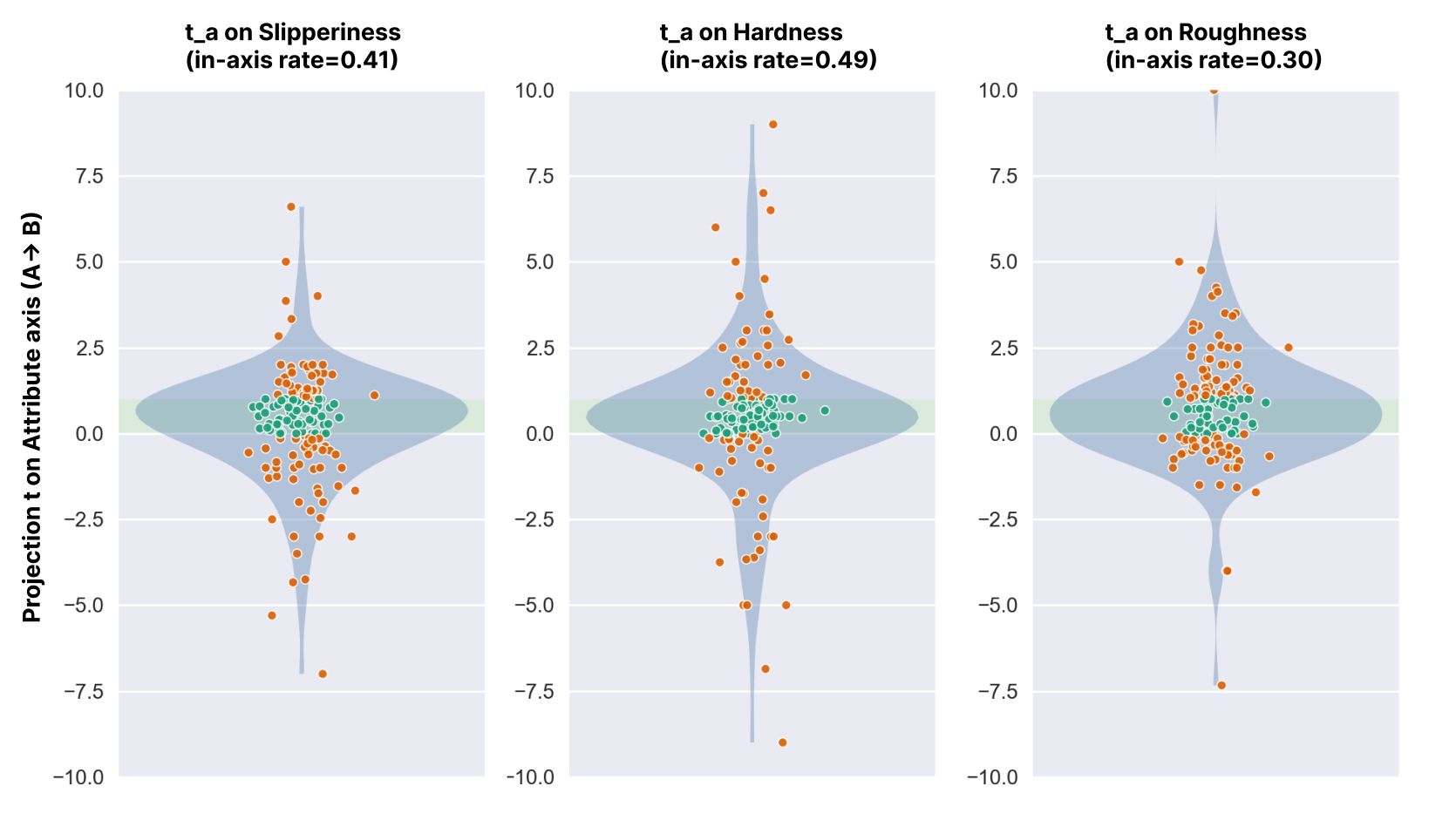}
  \caption{\textbf{Attribute-wise axis projections.} Each dot is a participant–pair instance. Green band denotes “between anchors” ($t_a\in[0,1]$). Inside-axis rates: \textbf{Slipperiness} $0.41$, \textbf{Hardness} $0.49$, \textbf{Roughness} $0.30$.}
  \label{fig:per-attr-t}
\end{figure}

We summarize observations with two visualizations:

\textbf{(1) Attribute-wise projection scatter }(Fig.~\ref{fig:per-attr-t})

For each participant, pair, and attribute $a\!\in\!\{R,S,H\}$, we compute the anchor–axis projection
\begin{equation}
t_a = \frac{(X_a - A_a)(B_a - A_a)}{(B_a - A_a)^2},
\label{eq:ta}
\end{equation}

\noindent which places the rating $X_a$ on a 1-D line from anchor $A_a$ to $B_a$: $t_a{=}0$ at $A$, $t_a{=}1$ at $B$, $t_a{<}0$ beyond $A$, and $t_a{>}1$ beyond $B$. Thus, $t_a\!\in[0,1]$ (“green band” in Fig.~\ref{fig:per-attr-t}) means the generated blend is rated between the two anchors along attribute $a$. A violin plot is fitted for each attribute subplot to show the distribution of the projected ratings.

\begin{figure*}
  \centering
  \includegraphics[width=\linewidth]{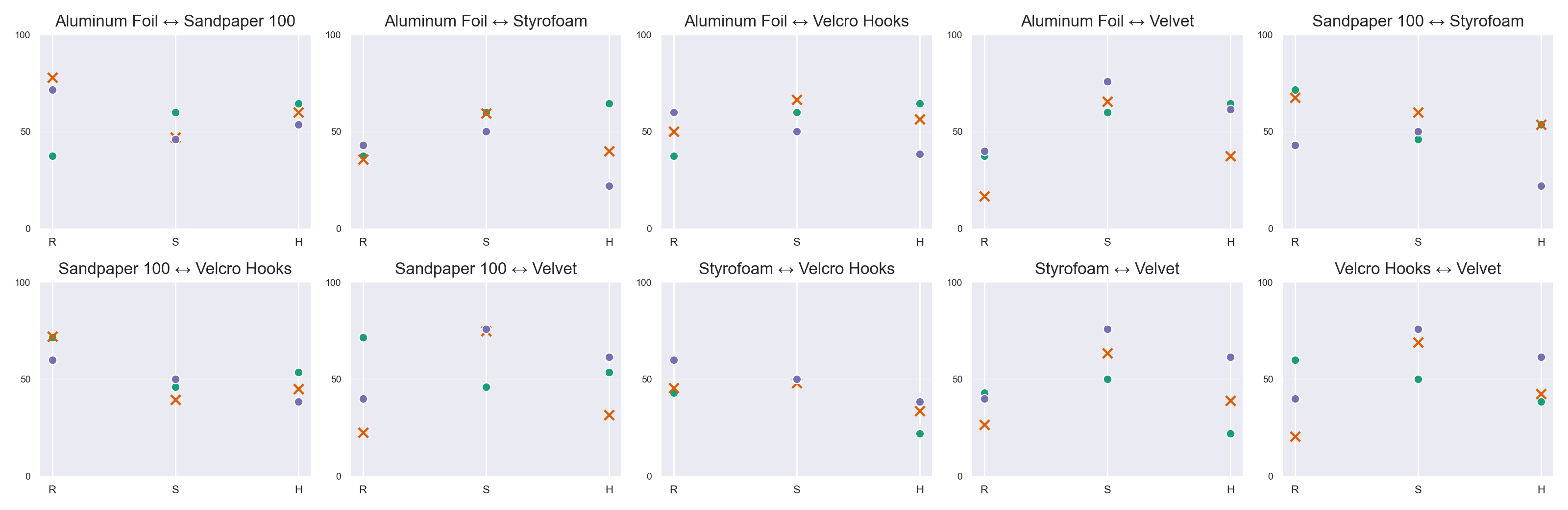}
  \caption{\textbf{Attribute-wise axis projections with distribution.} Violin plots (density) overlaid with per-trial scatter. The green band marks “between anchors” ($t_a\!\in[0,1]$).} %Inside-axis rates: \textbf{Hardness} $0.49$, \textbf{Slipperiness} $0.41$, \textbf{Roughness} $0.30$.}
  \label{fig:pairs-all-grid}
\end{figure*}

\textbf{(2) Per-pair median grid} (Fig.~\ref{fig:pairs-all-grid})  

For each pair, we take \emph{participant-wise medians} of the ratings for anchor $A$, the generated blend $X$, and anchor $B$ on each attribute and plot them together.

\paragraph*{Result}

\textbf{Hardness ($H$): strongest “between-anchors” placement.}
With an inside-axis rate of $0.49$, participants most often judged the generated hardness to sit between the two anchors (Fig.~\ref{fig:per-attr-t}, middle).
The violin distribution shows the tightest density within the green band, with relatively shallow tails, indicating frequent “between-anchors” placement and smaller excursions outside [0,1].
Pairs containing a compliant anchor (Velvet or Styrofoam) pulled $H$ \emph{below} an arithmetic midpoint, consistent with tap transients conveying onset stiffness while soft backings damp ringing (sub-additive firmness).
See \emph{Foil$\leftrightarrow$Styrofoam} and \emph{Sandpaper$\leftrightarrow$Velvet} in Fig.~\ref{fig:pairs-all-grid}.

\medskip
\textbf{Slipperiness ($S$): moderate “between-anchors,” biased by smooth films.}
Inside-axis rate is $0.41$ (Fig.~\ref{fig:per-attr-t}, left).
The violin plot shows a broader density with a slight skew toward higher $t_a$ values, corresponding to shifts toward smoother-film anchors (Foil or Velvet).
This indicates that participants tended to rate blends as \emph{more slippery} when a smooth surface film was present, largely independent of hardness.
This pattern aligns with our runtime friction estimate tied to latent neighbors and expressed as tangential damping (Eq.~\ref{eq:friction}).

\medskip
\textbf{Roughness ($R$): least “between-anchors,” rough anchor dominates.}
Inside-axis rate drops to $0.30$ (Fig.~\ref{fig:per-attr-t}, right).
The violin distribution exhibits the widest spread and heavy tails, consistent with the dominance of asperity cues—blends often inherit the rougher anchor’s grain strongly enough to exit the [0,1] interval.
Users frequently reported the blend retaining the properties of the rough anchor (Sandpaper or Hooks) even when paired with a smoother surface. It reflects that AR models emphasize high-band energy from micro-collisions. 
See \emph{Foil$\leftrightarrow$Hooks} and \emph{Sandpaper$\leftrightarrow$Velvet} in Fig.~\ref{fig:pairs-all-grid}.

\textbf{Putting \textit{R/S/H} together.}
Across pairs in Fig.~\ref{fig:pairs-all-grid}, users perceived the generated textures as \emph{plausible hybrids}: 
(i) $R$ is shaped by asperity statistics (rough anchor dominance), 
(ii) $H$ is governed by compliance (soft-backings reduce felt firmness), and 
(iii) $S$ follows surface films (smoothness over stiffness). 
These user ratings explain why combined analyses show “between-anchor” trends along the principal direction, yet exhibit bounded deviations when attributes pull differently, indicating coherent but \emph{non-collinear} cue mixing.

\subsection{Phase I: Qualitative Analysis (HXI)}
\label{sec:phase1_qual}
To complement anchor-referenced projections and per-pair median analyzes, we also assess interaction usefulness, engagement, perceived realism, cross-modal coherence, and mismatch, to provide a validity check on the overall experience.

We grounded our questionnaire in the Haptic Experience Inventory (HXI) factors—\textbf{Autotelics}, \textbf{Involvement}, \textbf{Realism}, \textbf{Harmony}, \textbf{Discord} (4 items each)—and adapted items to our usecase following~\cite{shi2025development}. Items used a 7-point Likert scale.

\begin{figure}[h]
  \centering
  \includegraphics[width=\linewidth]{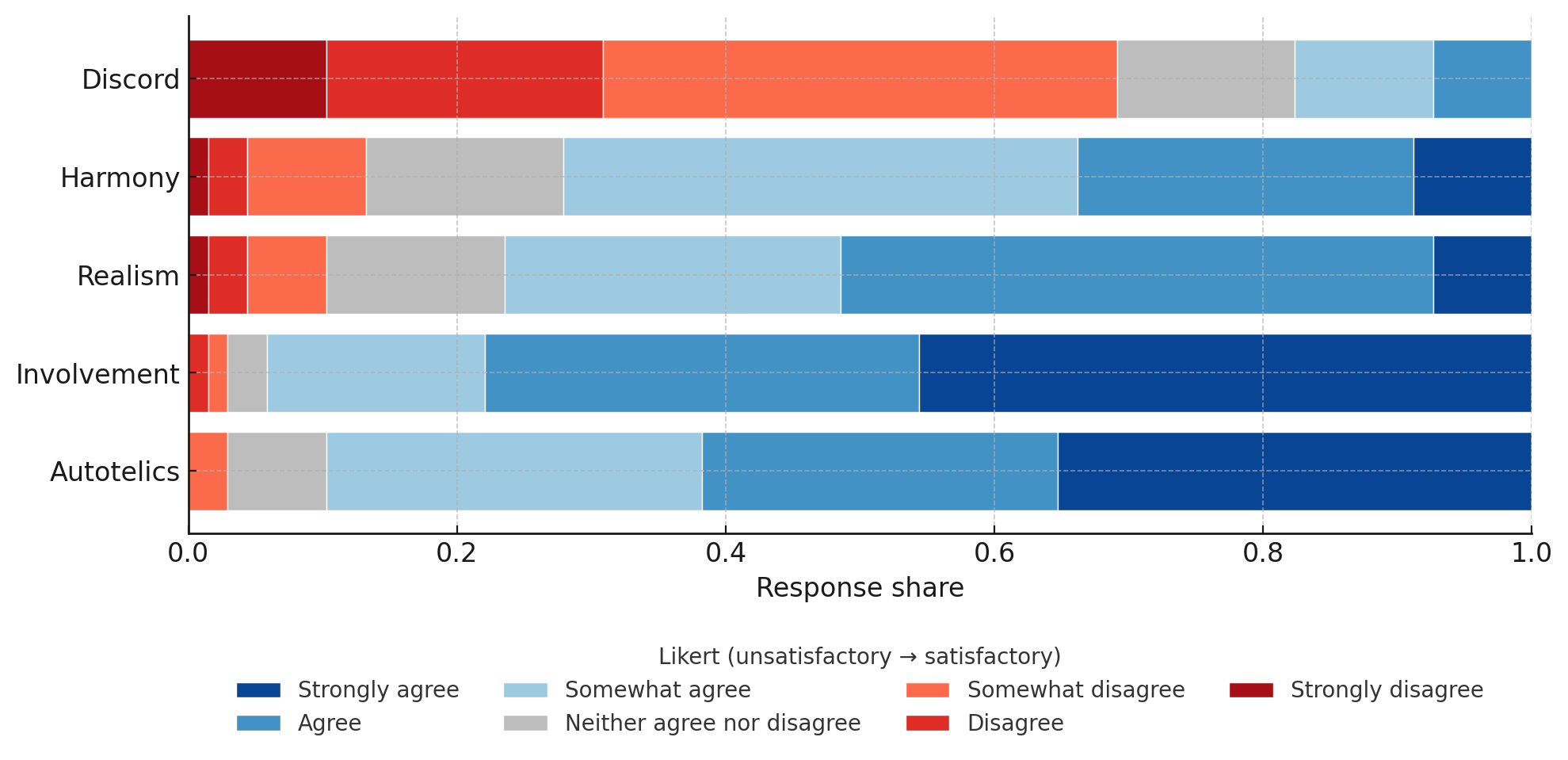}
  \caption{\textbf{HXI user ratings.} 100\% stacked distributions per factor (Discord reversed for scoring). Color encodes unsatisfactory$\rightarrow$satisfactory from red to blue}
  \label{fig:hxi-overview}
\end{figure}

HXI user ratings in Fig.~\ref{fig:hxi-overview} show three clear trends. First, \textbf{Autotelics} and \textbf{Involvement} are strongly right-skewed: participants \emph{enjoyed} the sensations and felt that the haptics \emph{helped them engage}. Second, \textbf{Realism} and \textbf{Harmony} lean to agreement: users experienced the outputs as \emph{plausible} and \emph{coordinated} with the concurrent image even though, at inference, image and haptics are only coupled via the \emph{shared prompt}. Third, reverse-scored \textbf{Discord} stays low (i.e., explicit reports of mismatch are rare).

Beyond HXI, we asked participants whether each interaction mode was useful and which attribute was easiest to judge. We visualize both as 100\% stacked bars with the same red$\rightarrow$blue ramp.

\begin{figure}[h]
  \centering
  \includegraphics[width=\linewidth]{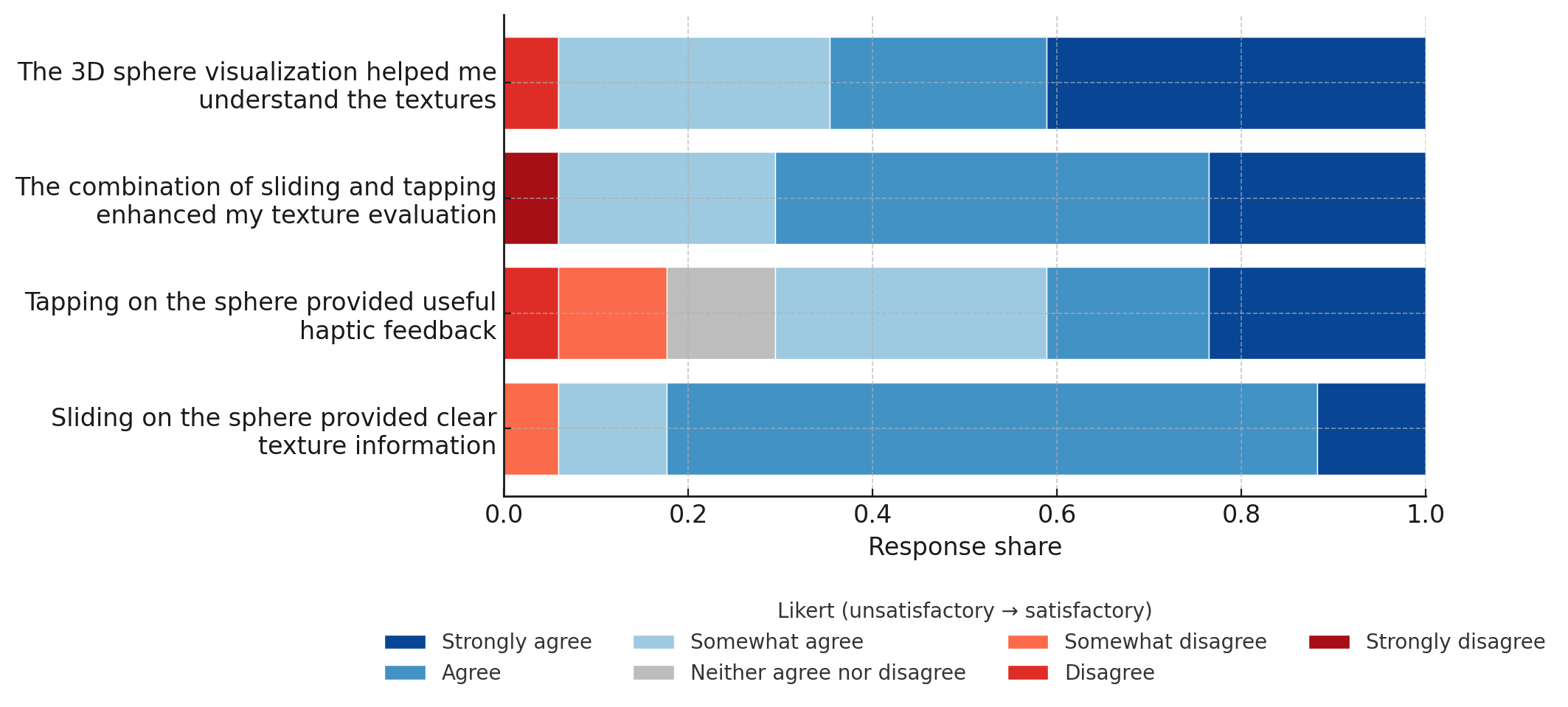}
  \caption{\textbf{Interaction usefulness} for Sliding, Tapping, their combination, and 3D visualization. Distributions concentrate toward the satisfactory (blue) end; the combination shifts furthest right, indicating complementarity rather than redundancy in haptic channels.}
  \label{fig:useful-stacked}
\end{figure}

\begin{figure}[h]
  \centering
  \includegraphics[width=\linewidth]{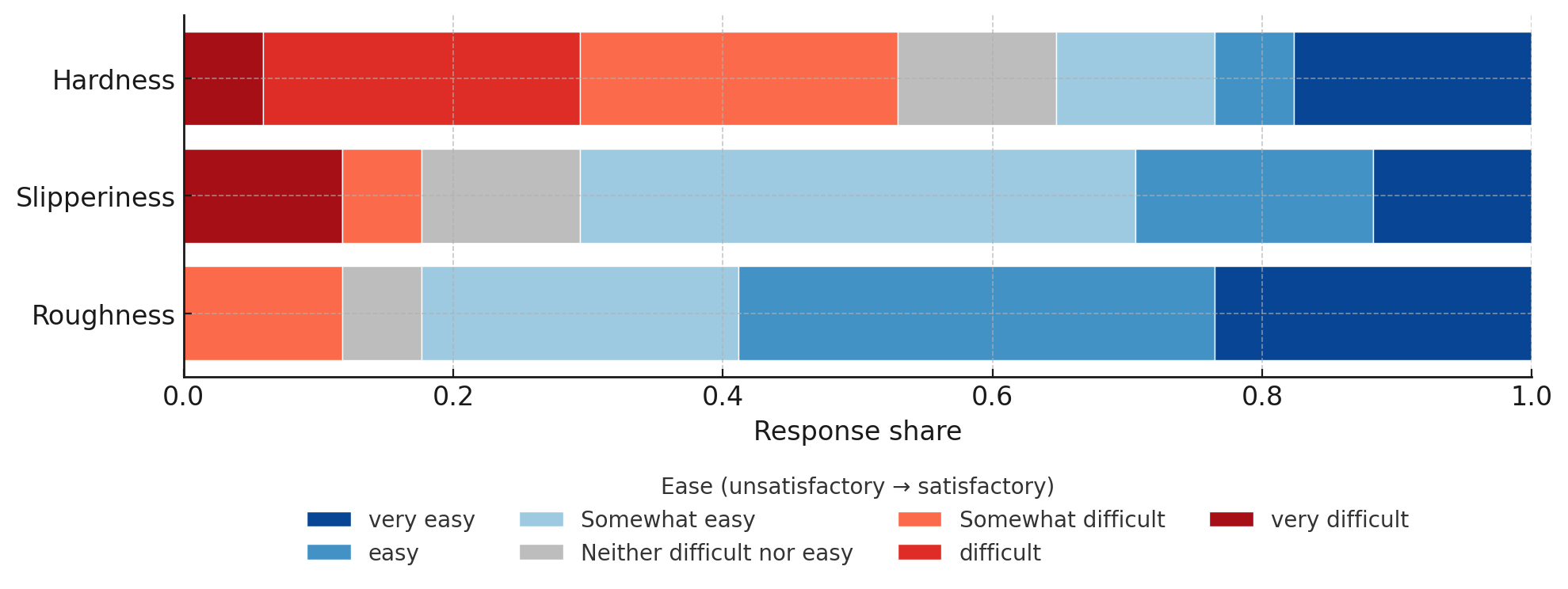}
  \caption{\textbf{Attribute-rating ease.} Roughness is easiest, Hardness next, Slipperiness hardest—consistent with our signal pathways (roughness salient while sliding; hardness from tap transients; slipperiness depends on friction/adhesion cues)}
  \label{fig:ease-stacked}
\end{figure}

For ratings of interaction usefulness in Fig.~\ref{fig:useful-stacked}, \emph{Sliding} concentrates responses on the satisfactory side, indicating that it gives clear texture information during exploration. \emph{Tapping} is also positive but with greater spread, matching its role in conveying short, high-contrast events (impact and compliance) that some users find subtler to interpret. Notably, the \emph{Sliding+Tapping} condition shifts the entire distribution further toward satisfaction, showing the two actions are complementary rather than redundant. The \emph{3D visualization} similarly skews toward satisfactory, supporting our choice to keep a visual reference even though the image is generated independently from the haptic latent.

In Fig.~\ref{fig:ease-stacked}, \textbf{Roughness} is rated the easiest attribute to judge, followed by \textbf{Hardness}, while \textbf{Slipperiness} is perceived as the most difficult. This ordering aligns with nature of the signals produced by our renderer: roughness is strongly conveyed in the sliding AR band, producing large and easily perceived variations; hardness is primarily carried by tap transients, which are distinct but brief; and slipperiness depends on friction that varies with contact force, speed, and micro-geometry, making it inherently noisier and more difficult to perceive reliably.

\subsection{Phase II: Language as Pragmatic Navigation}
\label{sec:phase2}

Each participant completed three \emph{prompt$\rightarrow$texture} trials. In each trial, they entered a short natural-language description of a texture they wanted to create. Participants explored the texture by tapping and sliding using the haptic device while viewing the image. After exploration, they rated their experience using the NASA–TLX~\cite{hart1988development} workload items and language-specific questions on realism, description-match, and satisfaction.

\begin{figure}[tb]
  \centering
  \includegraphics[width=\linewidth]{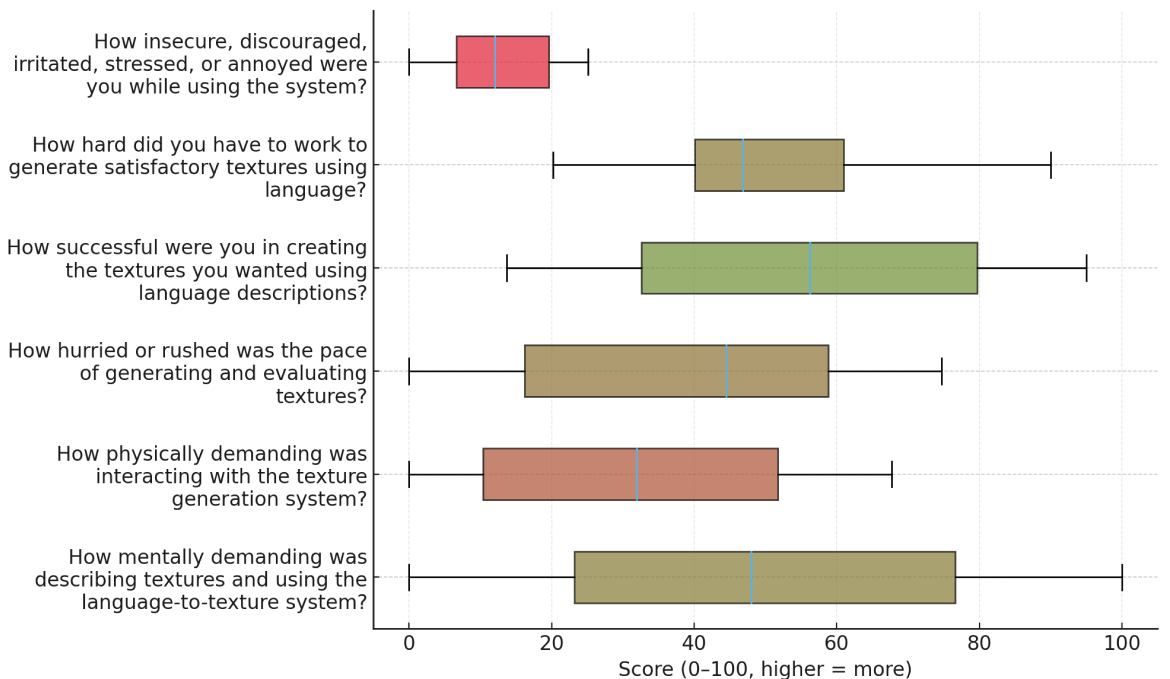}
  \caption{\textbf{NASA–TLX} Mental, physical, temporal demand, effort, frustration, and perceived success. Scores are shown on their native scales; lower is better for demand/effort/frustration, higher is better for success.}
  \label{fig:phase2-workload}
\end{figure}

\begin{figure}[tb]
  \centering
  \includegraphics[width=\linewidth]{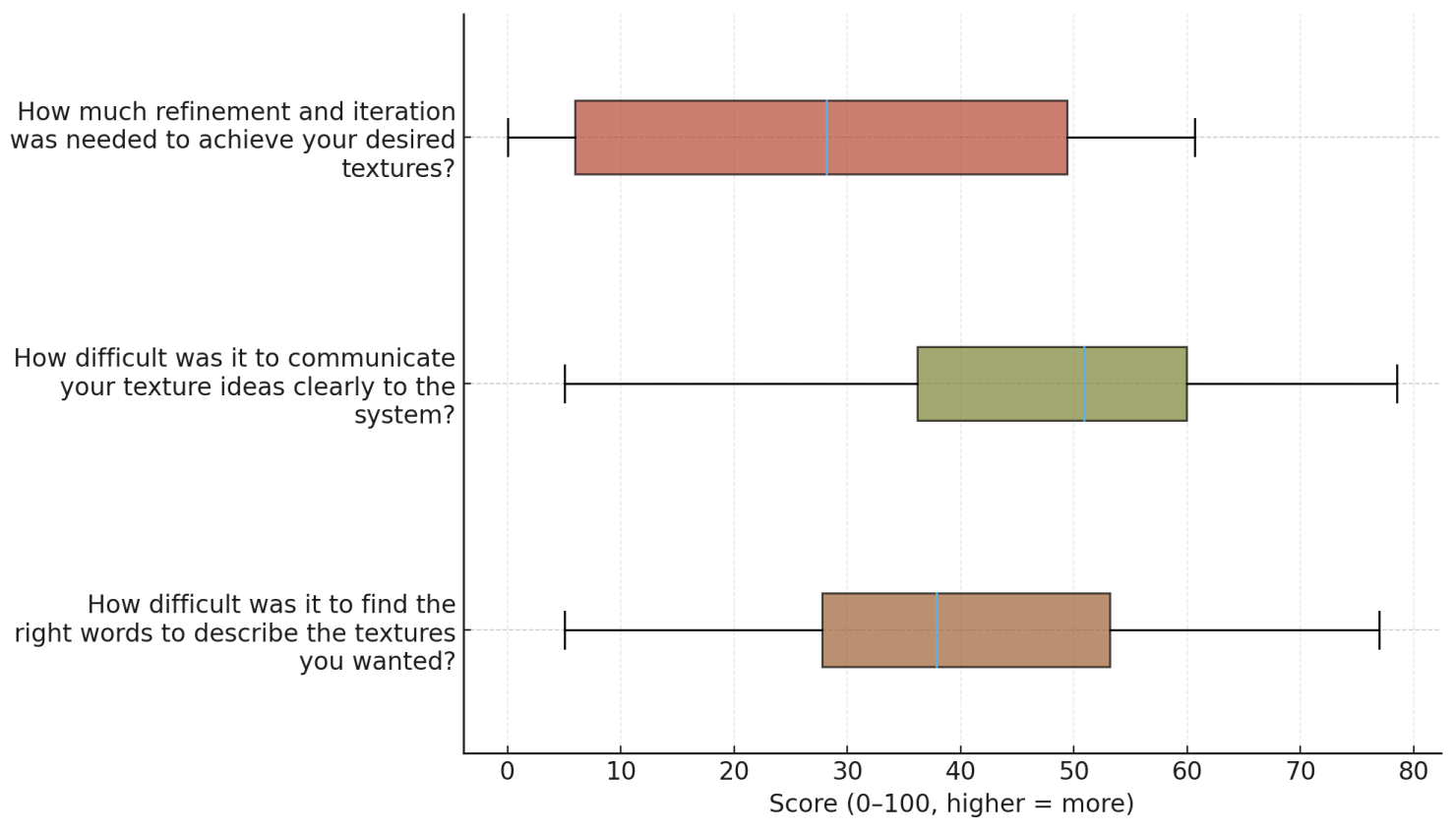}
  \caption{\textbf{Language-specific outcomes.} Agreement that the output matched the description and felt realistic, plus self-rated ability to create and refine using language prompts.}
  \label{fig:phase2-agree}
\end{figure}

\paragraph{Findings.}
As shown in Fig.~\ref{fig:phase2-workload}, satisfaction levels are consistently high across NASA–TLX dimensions. Mental and temporal demands convert to mid-to-high satisfaction, while physical effort remains lowest—expected for stylus-based interaction. Frustration scores are low, and NASA–TLX success is notably high, indicating that participants felt confident and effective during texture creation. Language-specific workload shows a similar trend, suggesting that phrasing prompts and making one single refinement felt straightforward and manageable.

Agreement ratings (Fig.~\ref{fig:phase2-agree}) also skew strongly toward satisfaction. Participants agreed that generated textures matched their intended descriptions and appeared realistic. Ratings for \emph{creation ability} were positive, while \emph{refinement ability} showed a small neutral portion but remained agreement-dominant overall. Together, these findings indicate that participants could use language as a \emph{pragmatic navigation tool} through the latent manifold; most users reached the texture they envisioned with one prompt and, at most, a light adjustment, all at moderate cognitive cost and minimal physical effort.

Open-ended feedback corroborates these findings. Participants frequently noted that the workflow “got me close” on the first try, and that a single adjective change (e.g., “less gritty,” “more waxy”) often produced the desired feel. Roughness and hardness were intuitive to control through adjectives such as “coarse,” “fine,” “soft,” or “rigid,” while slipperiness required more nuanced descriptors (“waxy,” “matte,” “oily”). Participants also valued the strong cross-modal coherence between visuals and haptics—“the image matched the feel”—and suggested lightweight refinements for future versions, including small attribute sliders for roughness, slipperiness, and hardness, a visible prompt history, A/B comparison thumbnails, and short hover hints linking adjectives to their expected haptic effects.

%% file: sections/6-Discussion.tex
\section{Discussion}
\label{sec:discussion}

\subsection{Language as a Control Modality for Texture Authoring}

\oldtext{Our findings demonstrate that natural language can serve as a structured and interpretable control interface for multimodal texture generation. Across the end-to-end examples and latent interpolations, linguistic prompts elicited consistent changes in both haptic and visual signals that align with human perceptual expectations. Participants were able to reason about how verbal modifiers such as soft, rough, or slippery translate into tactile outcomes, indicating that language-conditioned latents preserve meaningful perceptual gradients. The attribute-wise user ratings confirmed that generated intermediates were typically perceived as falling between reference anchors along physically interpretable dimensions rather than as arbitrary blends.}

\oldtext{Beyond signal coherence, the results highlight that users experienced the system as comprehensible and enjoyable to use. High ratings in the Haptic Experience Inventory (HXI) factors—particularly \emph{Autotelics}, \emph{Involvement}, and \emph{Realism}—show that participants found language-guided generation both engaging and believable. }

\oldtext{Together, these outcomes suggest that linguistic input not only controls model behavior effectively but also supports user intuition, making text a practical entry point for creative material design. In turn, this points toward hybrid authoring interfaces that integrate language prompts with light parametric controls, allowing designers to begin with a descriptive phrase and then refine specific perceptual attributes through direct manipulation.}

\newtext{Our findings validate natural language as an interpretable control for texture generation. Linguistic prompts reliably steered haptic and visual signals along perceptual gradients, with user ratings confirming that generated intermediates represent meaningful interpolations rather than arbitrary blends. High HXI scores (Autotelics, Involvement, Realism) further demonstrate that users found the workflow engaging and the results believable. Collectively, these outcomes suggest text is a practical, intuitive driver for creative design, paving the way for hybrid interfaces that combine high-level descriptive prompting with fine-grained parametric refinement.}

\newtext{We also examined the model's behavior on materials not present in the HaTT dataset. Because CLIP aligns unseen concepts (e.g., ``gelatin,'' ``skin'') with semantically related training examples in the latent space, the model generates haptics based on the closest learned physics. In our tests, ``skin'' consistently yielded sensations reported as realistic, and ``gel'' was described as ``soft and sticky''—likely interpolating from compliant materials like rubber. However, this extrapolation is bounded by the dataset's physical scope; ``sand,'' for instance, rendered as a rigid ``sandpaper-like'' texture, as the underlying training data lacks the granular mechanics of loose particles.}

\subsection{Limitations and threats to validity}
\begin{itemize}
  \item \textbf{Data coverage.} \oldtext{Training uses HaTT-derived AR/tap channels and our augmentations; there are no held-out \emph{zero-shot} materials. We therefore cannot claim open-set generalization. More diverse, multi-device datasets (including finger-surface recordings) are needed.}
\newtext{Current training relies on synthetic augmentation of the HaTT dataset. We are actively addressing this by compiling a corpus of $\sim$300 textures with sufficient intra-class variation to eliminate the need for augmentation. Additionally, the reliance on GPT-5 descriptions was necessitated by the unavailability of physical source samples for human evaluators; future work will utilize physically accessible libraries to enable human-verified haptic grounding.}
  \item \textbf{Device dependence.} Playback was on a 3-DOF kinesthetic device at 1\,kHz. Absolute feel (gain, bandwidth, friction rendering) may differ on other hardware; replication on alternative stylus/actuator stacks is a priority.
  \item \textbf{Decoupled visuals.} Images are generated by a separate diffusion model; while users rarely reported mismatch, we do not guarantee physical consistency (e.g., SVBRDF/roughness) beyond semantic agreement.
  \newtext{\item \textbf{Baseline Comparison.} Our evaluation focused on the proposed system in isolation. Since no other language-guided multimodal texture authoring tool currently exists, and manual parameter tuning represents a fundamentally different interaction paradigm, a direct baseline comparison was not feasible. Consequently, the reported metrics (e.g., HXI) should be interpreted as validation of the system's absolute usability and workflow feasibility, rather than evidence of quantitative improvement over existing methods.}
  \newtext{\item \textbf{Linguistic Robustness.} We did not systematically evaluate the model's sensitivity to paraphrasing. Future work will quantify latent space stability across synonymous descriptors (e.g., comparing ``smooth'' vs. ``polished'') to ensure that minor textual variations result in predictable, consistent haptic outputs.}
\end{itemize}

\subsection{Engineering lessons}
We observe three issues during system construction and user study: (i) participants report occasional prompt$\rightarrow$render latency spikes, (ii) default friction/roughness gains were sometimes too strong for some participants, and (iii) abrupt behavior occurred outside the recorded force--speed hull. Mitigations for (i) and (ii) include streaming decode and per-material gain auto-scaling. (iii) has since been resolved by incorporating \emph{edge models} that smoothly extend dynamics beyond the measured range, eliminating the discontinuities observed during early testing.

\subsection{Future work}

\textbf{Cross-modal alignment.} Our immediate goal is to make the diffusion image respond to, and constrain, the haptic latent rather than relying on text alone. We will (i) learn a shared, bidirectionally predictive space in which images and haptics encode to the same $z$ and can cross-reconstruct, and (ii) cross-condition the diffusion UNet on $z$ via a lightweight adapter (e.g., cross-attention) so that image microgeometry (grain/pores/finish) is consistent with AR band energy and tap rise/decay. To stabilize alignment, we will add weak physics-aware priors that correlate visual cues with haptic statistics and train on text–image–haptics triplets (paired or pseudo-paired) with consistency losses.

\textbf{Human Generated Texture.}
Our text-conditioning currently relies on AI-generated descriptions to represent material semantics. A valuable next step is to incorporate \textbf{human-annotated descriptions} of textures—collected through crowd-sourcing or expert labeling—to establish stronger grounding between language and human haptic perception. Such human supervision could refine semantic alignment from human tactile understanding.

\textbf{Beyond isotropy/rigidity.} We plan to relax our assumptions by modeling \emph{anisotropy} (directional grains, woven fabrics) and \emph{rate dependence} (viscoelastic media). Practically, this means collecting directional AR grids and variable-rate tap banks and extending the latent to $z(f,v,\theta,\dot v)$ so synthesis reflects exploration heading and speed changes. Additionally, modeling soft materials can be approached either through data driven or model-based methods.

\textbf{Authoring and portability.} For authoring, we will expose semantic sliders (rough$\leftrightarrow$smooth, hard$\leftrightarrow$soft, slippery$\leftrightarrow$sticky) that move $z$ along calibrated axes, add token-level attribution to show which words influenced each attribute, and provide certainty cues and auto-gain to avoid over/under-emphasis. For portability, we will release calibration recipes and learn device adapters so the same latent produces comparable feel across different end-effectors (e.g., voice-coil, piezo, ultrasonic friction).

%% file: sections/7-conclusion.tex
\section{Conclusion}
\oldtext{This work demonstrates that natural language can serve not only as a descriptive medium but as a functional control interface for texture authoring. By aligning language with a shared latent space spanning haptic and visual modalities, we show that prompts can evoke structured, physically meaningful changes in generated textures—ranging from surface compliance to frictional roughness—without manual parameter tuning. The resulting workflow allows users to navigate material space semantically rather than numerically, expressing design intent through words rather than low-level features.}

\oldtext{The findings point toward a broader paradigm for multimodal authoring: one where verbal descriptions act as anchors, and perceptual attributes—hardness, roughness, slipperiness—form an interpretable bridge between human language and machine generation. Our perceptual evaluation supports this vision, showing that language-guided outputs are experienced as coherent and controllable, while self-reports indicate high engagement and realism with minimal cognitive load.}

\oldtext{Still, realizing open-set, generalizable language–texture synthesis will require richer multimodal datasets, cross-device calibration, and explicit coupling between visual and haptic latents. As those resources mature, language may become a universal substrate for material interaction—making texture design accessible, adaptive, and deeply aligned with human perception.}

\newtext{
We presented a language-guided authoring system that synthesizes coordinated haptic (vibration, tapping) and visual textures from natural language. By aligning a multimodal VAE with CLIP, we established a semantic latent space where prompts like ``gritty stone'' yield physically grounded, cross-modally consistent signals. User evaluations confirmed that this latent structure correlates with human perception across roughness, hardness, and slipperiness, enabling an intuitive, prompt-driven workflow. Future work will address current data constraints by expanding the system to support bare-finger interactions and open-set material generalization.}